\documentclass[fleqn,usenatbib]{mnras}

\usepackage[T1]{fontenc}
\usepackage{ae,aecompl}
\usepackage[flushleft]{threeparttable}

\usepackage{graphicx}
\graphicspath{{figures/}}
\usepackage{amsmath,mathrsfs}
\usepackage{amssymb}
\usepackage{newtxtext,newtxmath}
\usepackage{mathtools}
\usepackage{pdflscape}	% Landscape pages
\usepackage{subcaption}
\usepackage{booktabs}
\usepackage[dvipsnames]{xcolor}
\usepackage{longtable}

\title[The dynamic magnetosphere of Swift J1818.0$-$1607]{The dynamic magnetosphere of Swift J1818.0$-$1607}

\author[M.~E.~Lower et al.]{\parbox{\textwidth}{
M.~E.~Lower,$^{1,2}$\thanks{E-mail: mlower@swin.edu.au}
S.~Johnston,$^{2}$
R.~M.~Shannon,$^{1,3}$
M.~Bailes,$^{1,3}$
and F.~Camilo$^{4}$
}
\\ \\
$^{1}$Centre for Astrophysics and Supercomputing, Swinburne University of Technology, PO Box 218, Hawthorn, VIC 3122, Australia\\
$^{2}$CSIRO Astronomy and Space Science, Australia Telescope National Facility, Epping, NSW 1710, Australia\\
$^{3}$OzGrav: The ARC Centre of Excellence for Gravitational-wave Discovery, Hawthorn VIC 3122, Australia\\
$^{4}$South African Radio Astronomy Observatory, 2 Fir Street, Black River Park, Observatory 7925, South Africa\\
}

\date{Accepted XXXX. Received YYYY; in original form ZZZZ}

\pubyear{0000}

\begin{document}
\label{firstpage}
\pagerange{\pageref{firstpage}--\pageref{lastpage}}
\maketitle

\begin{abstract}
    Radio-loud magnetars display a wide variety of radio-pulse phenomenology seldom seen among the population of rotation-powered pulsars.
    Spectropolarimetry of the radio pulses from these objects has the potential to place constraints on their magnetic topology and unveil clues about the magnetar radio emission mechanism. 
    Here we report on eight observations of the magnetar Swift J1818.0$-$1607 taken with the Parkes Ultra-Wideband Low receiver covering a wide frequency range from 0.7 to 4\,GHz over a period of 5 months.
    The magnetar exhibits significant temporal profile evolution over this period, including the emergence of a new profile component with an inverted spectrum, two distinct types of radio emission mode switching, detected during two separate observations, and the appearance and disappearance of multiple polarization modes.
    These various phenomena are likely a result of ongoing reconfiguration of the plasma content and electric currents within the magnetosphere.
    Geometric fits to the linearly polarized position angle indicate we are viewing the magnetar at an angle of $\sim$99$^{\circ}$ from the spin axis, and its magnetic and rotation axes are misaligned by $\sim$112$^{\circ}$. 
    While conducting these fits, we found the position angle swing had reversed direction on MJD 59062 compared to observations taken 15\,days earlier and 12\,days later.
    We speculate this phenomena may be evidence the radio emission from this magnetar originates from magnetic field lines associated with two co-located magnetic poles that are connected by a coronal loop. 
\end{abstract}


%%%%%%%%%%%%%%%%%%%%%%%%%%%%%%%%%%%%%%%%%%%%%%%%%%

%%%%%%%%%%%%%%%%% BODY OF PAPER %%%%%%%%%%%%%%%%%%

%%%%%%%%%%%%%%%%%%%%%%%%%%%%%%%%%%%%%%%%%%%%%%%%%%
\section{Introduction}

\begin{table*}
\begin{center}
\caption{Parkes UWL observations of Swift J1818.0$-$1607, along with the number of recorded profile components and rotation measures from {\sc rmfit} and direct Stokes $Q$-$U$ fits.\label{tbl:obs}}
\renewcommand{\arraystretch}{1.2}
\setlength{\tabcolsep}{4pt}
\begin{tabular}{lccccccc}
\hline
Observation & MJD & Frequency & Bandwidth & Length & No. profile & RM ({\sc rmfit}) & RM ($Q$-$U$) \\
(UTC)       &     & (MHz)     & (MHz)     & (s)    & components  & rad\,m$^{-2}$    & rad\,m$^{-2}$    \\
\hline
2020-05-08-18:14:52 & 58977 & 2368 & 3328 & 639  & 1 & $1440.48 \pm 0.09$ & $1440.1 \pm 0.8$ \\
2020-06-09-11:34:36 & 59009 & 2368 & 3328 & 616  & 2 & $1440.86 \pm 0.04$ & $1441.7 \pm 0.7$ \\
2020-07-17-09:35:10 & 59047 & 2368 & 3328 & 616  & 2 & $1441.72 \pm 0.05$ & $1439.2^{+0.3}_{-0.2}$ \\
2020-08-01-13:19:01 & 59062 & 2368 & 3328 & 2440 & 2 & $1441.72 \pm 0.05$ & $1447.4 \pm 0.4$ \\
2020-08-13-10:37:18 & 59074 & 2368 & 3328 & 617  & 2 & $1439.05 \pm 0.07$ & $1439 \pm 2$ \\
2020-08-26-09:06:53 & 59087 & 2368 & 3328 & 623  & 3 & $1439.47 \pm 0.04$ & $1440.3 \pm 0.4$ \\
2020-09-17-04:40:32 & 59109 & 2368 & 3328 & 1139 & 2 & $1443.53 \pm 0.04$ & $1440.5 \pm 0.4$ \\
2020-10-06-07:06:18 & 59128 & 2368 & 3328 & 618  & 2 & $1445.79 \pm 0.08$ & $1444.2 \pm 0.6$ \\
\hline
\end{tabular}
\renewcommand{\arraystretch}{}
\end{center}
\end{table*}

Swift J1818.0$-$1607 belongs to a sub-class of slowly rotating, young neutron stars that possess unusually high X-ray and gamma-ray luminosities, commonly referred to as magnetars.
They are believed to be powered by the dissipation of their ultra-strong internal magnetic fields as opposed to the slow release of stored angular momentum \citep{Thompson1995}.
Most are detected as persistent sources of high energy electromagnetic radiation and occasionally undergo periods of high activity, where bursts of intense X-ray and gamma-ray emission are commonplace.
If a magnetar was born rapidly rotating, its internal magnetic field will be strongly wound up \citep{Duncan1992}.
Relaxation of the internal magnetic field exerts strong magnetic forces on the crust that can lead to local or even global twists in the magnetic field due to horizontal plastic deformation or fracturing of the crust (i.e a starquake) if these stresses are allowed to build up over time \citep{Thompson2002}.
It is the sudden twisting of the magnetic field lines along with magnetic re-connection events that are believed to power magnetar outbursts \citep[see][for a review]{Kaspi2017}.

\citet{Beloborodov2009} showed the current bundles that flow along a twist near the dipole axis of the magnetosphere can generate the conditions required for coherent radio emission to take place, potentially explaining why a handful of active magnetars have now been detected as radio pulsars.
These `radio-loud' magnetars exhibit an extremely diverse variety of radio emission phenomenology that are rarely displayed by less magnetic rotation-powered pulsars. 
Both their average and single pulse profiles have high degrees of linear polarization, typically in excess of 90\,per cent \citep{Kramer2007, Camilo2007a, Levin2010, Eatough2013}, and often possess extremely flat radio spectra \citep{Levin2012, Torne2015, Dai2019}.
The untwisting of their dynamic magnetic fields and associated electric currents following an outburst are imprinted in their radio profiles, which show variations in intensity and polarization, along with the emergence or disappearance of profile components on timescales ranging from a few hours to many months \citep[e.g.][]{Camilo2007d, Camilo2016, Scholz2017}.
The sweep of the linear polarization position angle can be interpreted geometrically, as has been done for several of the radio magnetars \citep{Camilo2007b, Camilo2007a, Levin2012}. However, deviations from the standard models often employed to fit the position angle swing have led some to speculate on the role of emission from closed magnetic field lines and contributions from higher-order multipole magnetic fields \citep[e.g.][]{Kramer2007}.
Despite these deviations from standard pulsar behaviour, radio-loud magnetars generally have higher spin-down luminosities than most `radio-quiet' magnetars, potentially pointing to a strong relationship with young radio pulsars (see \citealt{Rea2012} and discussions therein).

Unlike standard radio pulsars, the single pulses detected from magnetars are typically comprised of many `spiky' sub-pulses that show highly variability in intensity and width on a pulse-to-pulse basis \citep{Serylak2009, Levin2012, Pearlman2018}.
Similarities between the single pulse properties of magnetars and the phenomenology of fast radio bursts \citep[FRBs; e.g.][]{Pearlman2018, Maan2019}, combined with numerous FRB progenitor theories that invoke a magnetar central engine tentatively indicate radio magnetars within the Milky-Way may be galactic analogues to FRB progenitors \citep[e.g.][]{Wadiasingh2019}.
This possible connection has been strengthened by the detection of an extremely luminous, millisecond-duration radio burst from SGR 1935$+$2154 by the CHIME/FRB and STARE2 experiments \citep{CHIME2020, STARE2020}.

With a spin-period of approximately $1.4$\,s, Swift J1818.0$-$1607 is among the fastest rotating pulsars that show magnetar-like activity.
A secular spin-down rate of $4.6 \times 10^{-11}$\,s\,s$^{-1}$ and an inferred surface dipole magnetic field strength of $2.5 \times 10^{14}$\,G \citep{Champion2020}, place Swift J1818.0$-$1607 among the growing population of known galactic magnetars \citep{Olausen2014}\footnote{\url{http://www.physics.mcgill.ca/~pulsar/magnetar/main.html}}.
Observations by the Effelsberg and Lovell radio telescopes soon after its discovery revealed the magnetar to be radio-bright \citep{Champion2020}, making it only the fifth radio-loud magnetar.

Like other radio magnetars, its single pulses are comprised of narrow, spiky sub-pulses \citep[see Figure 3 of][]{Esposito2020}, with a high degree of linear polarization across a wide range of frequencies \citep{Lower2020}.
However, its unusually steep radio spectrum and lower than anticipated quiescent X-ray luminosity \citep{Esposito2020} seem to imply it shares more in common with more ordinary rotation-powered pulsars than other radio-loud magnetars.
These irregular properties and similar behaviour to that of PSR J1119$-$6127 following its 2016 outburst \citep{Archibald2016, Dai2018} led to speculation that Swift J1818.0$-$1607 may represent a possible missing link between magnetars and the population of magnetar-like, high magnetic field strength (high B-field) pulsars \citep{Hu2020}.

In this work, we explore the spectral, temporal and polarimetric properties of Swift J1818.0$-$1607 across the 3.3\,GHz bandwidth of the Ultra-Wideband Low (UWL) receiver system of the CSIRO Parkes 64-m radio telescope (also known as \textit{Murriyang}), covering eight epochs after its discovery in March 2020 until October 2020.
The details of our observations along with the calibration and data cleaning strategies are summarised in Section~\ref{sec:obs}.
Analyses of the magnetars profile and spectral evolution, the discovery of two kinds of discrete emission mode changing at two different epochs and geometric analyses based on fits to the linearly polarized position angle are presented in Sections~\ref{sec:spectra} through~\ref{sec:pa_swing}.
The implications of our analyses and results are discussed in Section~\ref{sec:disc}, with a particular focus on potential physical models that may describe the apparent variations in viewing geometry and polarized emission.
We also relate our observations to the transient behaviour of other radio magnetars and high B-field pulsars. 
A summary of our findings along with concluding remarks are presented in Section~\ref{sec:conc}.

%%%%%%%%%%%%%%%%%%%%%%%%%%%%%%%%%%%%%%%%%%%%%%%%%%
%%%%%%%%%%%%%%%%%%%%%%%%%%%%%%%%%%%%%%%%%%%%%%%%%%
\section{Observations}\label{sec:obs}

\begin{figure*}
    \centering 
    \includegraphics[width=\linewidth]{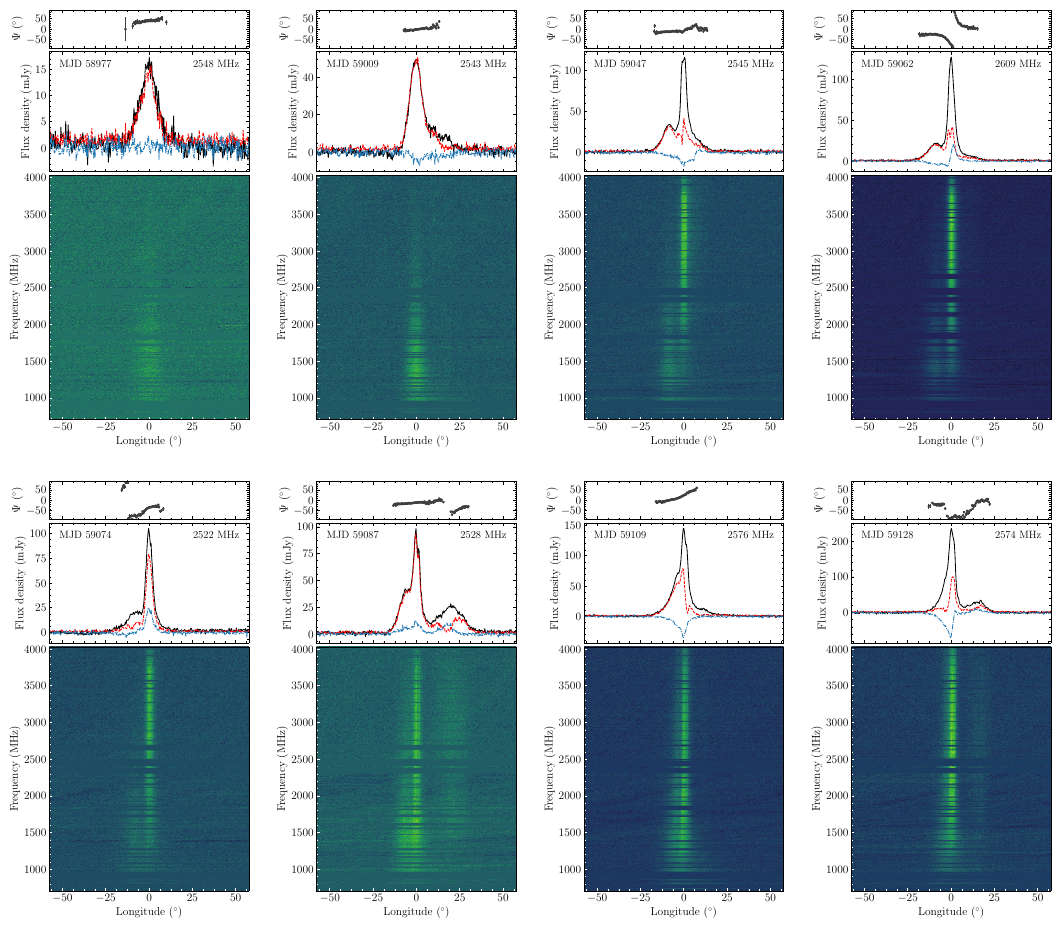}
    \caption{Parkes UWL observations of Swift J1818.0$-$1607. Each plot depicts the linear polarization position angle ($\Psi$) (top panel), polarization profile with total intensity in black, linear polarization in red and circular polarization in blue (middle panel), and the phase resolved total intensity spectrum (bottom panel).}
    \label{fig:obs}
\end{figure*}

Following its discovery in March 2020, we began a regular monitoring campaign of Swift J1818.0$-$1607 with the Parkes UWL receiver system \citep{Hobbs2020} under the P885 project (PI: F.~Camilo). 
During each observation we typically recorded $\sim$10-minutes of full Stokes pulsar search-mode data covering the full 3328\,MHz bandwidth of the UWL with 1\,MHz channels and 128\,$\mu$s sampling via the {\sc medusa} backend, where each frequency channel was coherently dedispersed with a dispersion measure (DM) of 706\,pc\,cm$^{-3}$. 
We created {\sc psrfits} \citep{Hotan2004} format archives with 1024\,phase bins by folding the psrfits-format search-mode data at the topocentric pulse period of the magnetar via {\sc dspsr} \citep{vanStraten2011}.
Calibration and cleaning of the data were performed via the methodology outlined in \citet{Lower2020}.
We note for the two observations performed on MJD 58977 and MJD 59009, we used noise diode scans taken 20\,minutes after and 30\,minutes before the respective Swift J1818.0$-$1607 observations on these dates.
User error prevented the noise diode from activating during the originally scheduled scans.
Later observations were not affected by this issue.
We tested for inconsistencies in the calibration by measuring the rotation measure (RM) of the polarization spectra at each epoch using both the brute-force method implemented in the {\sc rmfit} tool of {\sc PSRCHIVE} (searched over RM values between $-2000$ and $2000$\,rad\,m$^{-2}$ with 4000 steps), and a {\sc Python} implementation of the direct Stokes $Q$ and $U$ fitting technique described in \citet{Bannister2019}.
The resulting RM measurements, along with details of each observation are presented in Table~\ref{tbl:obs}.
Note the uncertainties of the {\sc rmfit} values are clearly underestimated by about an order of magnitude when compared to those obtained from the $Q$-$U$ spectral fits. 
While our recovered RM values deviate from the previously reported value of $1442.0 \pm 0.2$\,rad\,m$^{-2}$ \citep{Lower2020}, they are consistent with expected variations due to propagation through the ionosphere at the location of Parkes \citep[e.g.][]{Han2018}. 
Following this test, we applied the nominal RM of $1442.0$\,rad\,m$^{-2}$ referenced to the central observing frequency of 2368\,MHz to each observation.
As our observations are too sparse for a phase connected timing solution to be obtained, all profiles that we show in Figure \ref{fig:obs} were manually aligned so the total intensity maximum is located at a pulse longitude of $0^{\circ}$. 

%%%%%%%%%%%%%%%%%%%%%%%%%%%%%%%%%%%%%%%%%%%%%%%%%%
%%%%%%%%%%%%%%%%%%%%%%%%%%%%%%%%%%%%%%%%%%%%%%%%%%
\section{Profile and spectral evolution}\label{sec:spectra}

We show the polarization profiles, linear polarization position angle (PA; $\Psi$) swings and phase-resolved total intensity spectra for all eight observations in Figure~\ref{fig:obs}.
The emission profiles on MJD 58978 and 59009 are similar to the profile presented in \citet{Lower2020} and the subset of those in \citet{Champion2020} where a single, highly linearly polarized component with a steep spectrum and flat PA were detected. 
None of the averaged profiles shows evidence for the second component that was occasionally observed by \citet{Champion2020}.
However, we later show that a handful of pulses from this previously reported secondary component were detected on MJD 59009.
A new profile feature with an inverted spectrum emerged between MJD 59009 and 59047.
Reports from other facilities suggest the emission from this profile component is detectable up to frequencies as high as 154\,GHz \citep[e.g.][]{Torne2020ATel}.
This inverted-spectrum component persists throughout our later observations.
On the other hand, the steep-spectrum component gradually weakens and appears evolve toward more positive values of pulse longitude.
By MJD 59128 it is almost completely overlaps with the inverted-spectrum component.
Similar longitudinal evolution of individual profile components was detected in the pulsed radio emission of XTE J1810$-$197 following its 2018 outburst \citep{Levin2019}.
A third, weaker component that possesses a flat spectrum was detected on MJD 59087 and again as the secondary component on MJD 59128. 
The 90-degree jump in the PA along with the dip in linear polarization of this component on MJD 59087 are indicative of an orthogonal polarization mode (OPM), as opposed to the PA offset in the secondary component in Figure 7 of \citet{Champion2020}.
An OPM is also clearly visible in the leading profile component detected on MJD 59128.

The spectrum of the magnetar has evolved significantly since it was first detected in March 2020.
While comparisons of the phase-resolved spectral index would be preferable, each of the multi-component profiles exhibit variable spectral indices, hence the effects of interstellar scattering would bias our results towards spectra with low-frequency turnovers as the radio flux at low frequencies becomes increasingly spread out as a function pulse longitude.
As a result, we were limited to computing the phase-averaged spectral index at each epoch.
First, we split the UWL band into thirteen 256\,MHz-wide subbands that were then averaged in time and frequency to create a one-dimensional pulse profile for each subband.
Corrections to the profile baseline were performed using {\sc PSRCHIVE}.
We then computed the continuum flux density at each subband by averaging over the on-pulse region of each profile as
\begin{equation}
    S_{\nu} = \frac{1}{N_{\rm bin}} \sum_{i}^{N_{\rm on}} S_{\nu,i},
\end{equation}
where $N_{\rm bin}$ is the total number of phase bins, $N_{\rm on}$ is the number of phase bins covered by the on-pulse region and $S_{\nu,i}$ is the flux at the $i$-th phase bin. 
We set the on-pulse window to be between $\phi = -18^{\circ}$ to $29^{\circ}$ for sub-bands above 1.5\,GHz, and extend to $\phi = 90^{\circ}$ below 1.5\,GHz in order to account for scatter broadening. For the MJD 59087 and 59128 observations, the extended on-pulse window was used for the full band to accommodate the additional profile components. 
The flux uncertainty is computed from the normalised root-mean-square (RMS) of the off-pulse region as
\begin{equation}\label{eqn:flux_err}
    \sigma_{S,\nu} = \frac{\sqrt{N_{\rm on}}}{N_{\rm bin}} \sqrt{\sum_{i}^{N_{\rm off}} S_{\nu,i}^{2}},
\end{equation}
where $N_{\rm off}$ is the number of bins covering the off-pulse region.
The resulting flux density measurements can be found in the supplementary materials.
We then fit the resulting flux density spectra using either a simple power-law function
\begin{equation}
    S(x) = a\,x^{\kappa},
\end{equation}
where $a$ is a scaling parameter, $x = \frac{\nu}{\text{1\,GHz}}$ and $\kappa$ is the spectral index, or a broken power-law of the form
\begin{equation}
    S(x) = a
    \begin{cases} 
      x^{\kappa_{1}} & \text{if}\, \nu \leq \nu_{b} \\
      x^{\kappa_{2}}x_{b}^{\kappa_{1} - \kappa_{2}} & \text{otherwise}
   \end{cases},
\end{equation}
where $x_{b} = \frac{\nu_{b}}{\text{1 GHz}}$, $\nu_{b}$ is the frequency of the spectral break and $\kappa_{1}$ and $\kappa_{2}$ are the respective spectral indices before and after the spectral break.
Posterior distributions for the spectral parameters were sampled using {\sc Bilby} \citep{Ashton2019} as a wrapper for the {\sc dynesty} nested sampling algorithm \citep{Speagle2019}.
We assumed a Gaussian likelihood function of the form
\begin{equation}\label{eqn:likelihood}
    \mathcal{L}(d | \theta) = \prod_{i}^{N} \frac{1}{\sqrt{2\pi\sigma}} \exp \Big[ -\frac{(d_{i} - \mu_{i}(\theta))^{2}}{2\sigma^{2}} \Big],
\end{equation}
where $N = 13$ is the number of frequency subbands, $d$ is the measured flux density, $\mu(\theta)$ is the spectrum model described by parameters $\theta$ and $\sigma^{2} = \sigma_{S,\nu}^{2} + \sigma_{Q}^{2}$ is the uncertainty in the flux densities added in quadrature with an additional error parameter ($\sigma_{Q}$) to account for any systematic errors not accounted for in Equation~\ref{eqn:flux_err}.
We also assumed uniform priors between $-10$ and $10$ for the spectral indices, and a uniform prior spanning 700\,MHz to 4000\,MHz for the spectral break.

\begin{table}
\begin{center}
\caption{Results from spectral fits and associated log Bayes factors. Observations with only a single spectral index listed are those best described by a simple power-law. Those with two are best fit by a broken power-law.\label{tbl:spectral}}
\renewcommand{\arraystretch}{1.2}
\setlength{\tabcolsep}{4pt}
\begin{tabular}{lcccc}
\hline
MJD & $\ln(\mathcal{B}^{\rm BPL}_{\rm SPL})$ & $\kappa_{1}$ & $\kappa_{2}$ & $\nu_{b}$ (MHz) \\
\hline
58977 & $0.7$ & $-1.7^{+0.2}_{-0.3}$ & $-$ & $-$ \\
59009 & $0.1$ & $-2.7 \pm 0.1$ & $-$ & $-$ \\
59047 & $9.0$ & $-2.0 \pm 0.1$ & $0.4 \pm 0.2$ & $1801^{+111}_{-88}$ \\
59062 & $8.0$ & $-1.9 \pm 0.2$ & $0.4 \pm 0.2$ & $1693^{+105}_{-126}$ \\
59074 & $1.5$ & $-1.2^{+0.2}_{-0.3}$ & $0.2^{+0.5}_{-0.4}$ & $2034^{+409}_{-418}$ \\
59087 & $-0.6$ & $-1.2 \pm 0.2$ & $-$ & $-$ \\
59109 & $-0.8$ & $-1.0 \pm 0.1$ & $-$ & $-$ \\
59128 & $-1.3$ & $-0.5 \pm 0.1$ & $-$ & $-$ \\
\hline
\end{tabular}
\renewcommand{\arraystretch}{}
\end{center}
\end{table}

We employed Bayesian model selection to determine which spectral model best described the data. 
The resulting Bayes factors along with the median recovered values (and associated 68\,per cent confidence intervals) for the preferred spectral models are presented in Table~\ref{tbl:spectral}.
Our measurements for the single component profiles, in addition to the values of $\kappa_{1}$ on MJD 59047 and 59062, are consistent with the spread of spectral indices between $-3.6$ and $-1.8$ presented in \citet{Champion2020}.
The recovered values of $\kappa_{2}$ are consistent with the magnetar spectrum being inverted or close to flat at frequencies above 1.6 to 2.0\,GHz.
Consistently flat spectra were also obtained by observations of Swift J1818.0$-$1607 by the Deep Space Network between 2.3 and 8.4\,GHz on MJD 59045 by \citep{Majid2020ATelB}, who obtained a spectral index of $0.3 \pm 0.2$. 

The phase averaged spectrum on MJD 59087 and beyond are best described by the single power-law model, each showing a significant amount of flattening when compared to the previous observations.
Although the spectral index of $-1.2 \pm 0.2$ on MJD 59087 is consistent with the value of $\kappa_{1}$ measured on MJD 59074, it appears to have transition back to a single power-law spectrum, albeit one that is much flatter than seen in earlier observations.
This can be attributed to a combination of averaging over the additional flat spectrum components detected on MJD 59087 and 59128, and the apparent weakening and increasing level of overlap between the steep- and inverted-spectrum components that we mentioned earlier.

%%%%%%%%%%%%%%%%%%%%%%%%%%%%%%%%%%%%%%%%%%%%%%%%%% 
%%%%%%%%%%%%%%%%%%%%%%%%%%%%%%%%%%%%%%%%%%%%%%%%%%
\section{Emission mode switching}\label{sec:modes}

At least two magnetars show evidence for their radio emission switching between multiple, quasi-stable radio profiles (mode-changing) or between an `on' and `off' state (nulling).
\citet{Camilo2007a} and \citet{Halpern2008} reported at least two types of discrete state-changes in the single-pulse behaviour of 1E~1547.0$-$5408, while \citet{Yan2018} noted the Galactic Centre magnetar SGR 1745$-$2900 would randomly switch between two emission modes in addition to exhibiting nulling.
Sudden changes in the profile shape of XTE J1810$-$197 were also reported by \citet{Camilo2007b} approximately once every 15 hours, while the polarization properties of PSR J1622$-$4950 could be categorised into four different sub-classes \citep{Levin2012}.
However it is unclear if the phenomena in the latter two magnetars were genuine mode changes or not.
Both mode-changing and nulling are thought to be related to same phenomena: variations in (or a complete failure of) the coherent radiation mechanism due to large-scale redistribution of current flows and plasma content in the pulsar magnetosphere \citep{Kramer2006, Wang2007, Timokhin2010}.
The resulting changes in particle outflows and the associated torque acting to slow the neutron star spin over time have previously been linked to correlated profile shape and spin-down variations in a number of pulsars \citep{Lyne2010}. 

\begin{figure}
    \centering 
    \includegraphics[width=\linewidth]{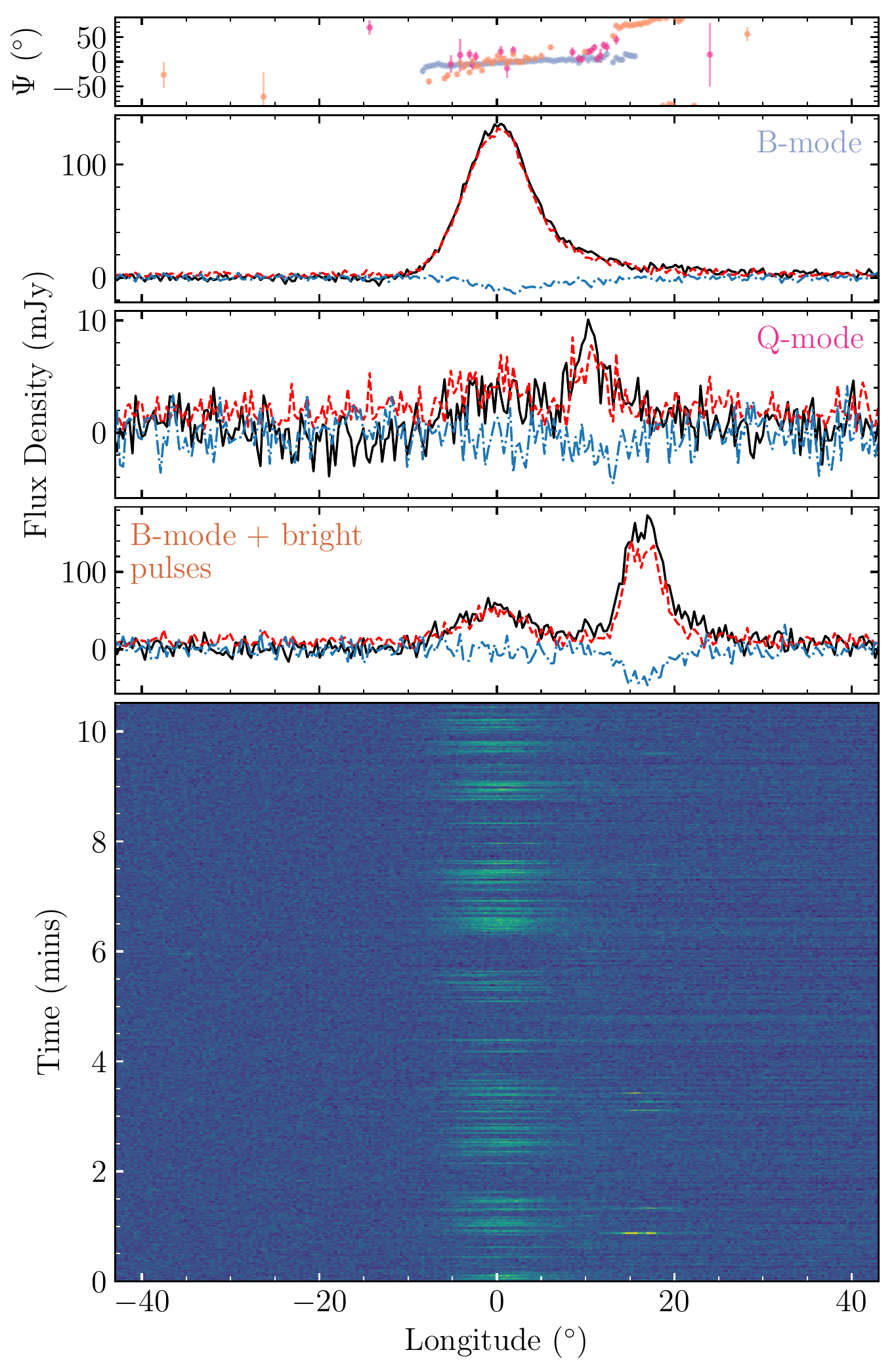}
    \caption{Top: comparison of the polarization profiles for the two emission modes detected on MJD 59009 and the profile after averaging only sub-integrations containing bright pulses from the second component. Bottom: stack of single pulses.}
    \label{fig:mode_switch_1}
\end{figure}

\begin{figure}
    \centering 
    \includegraphics[width=\linewidth]{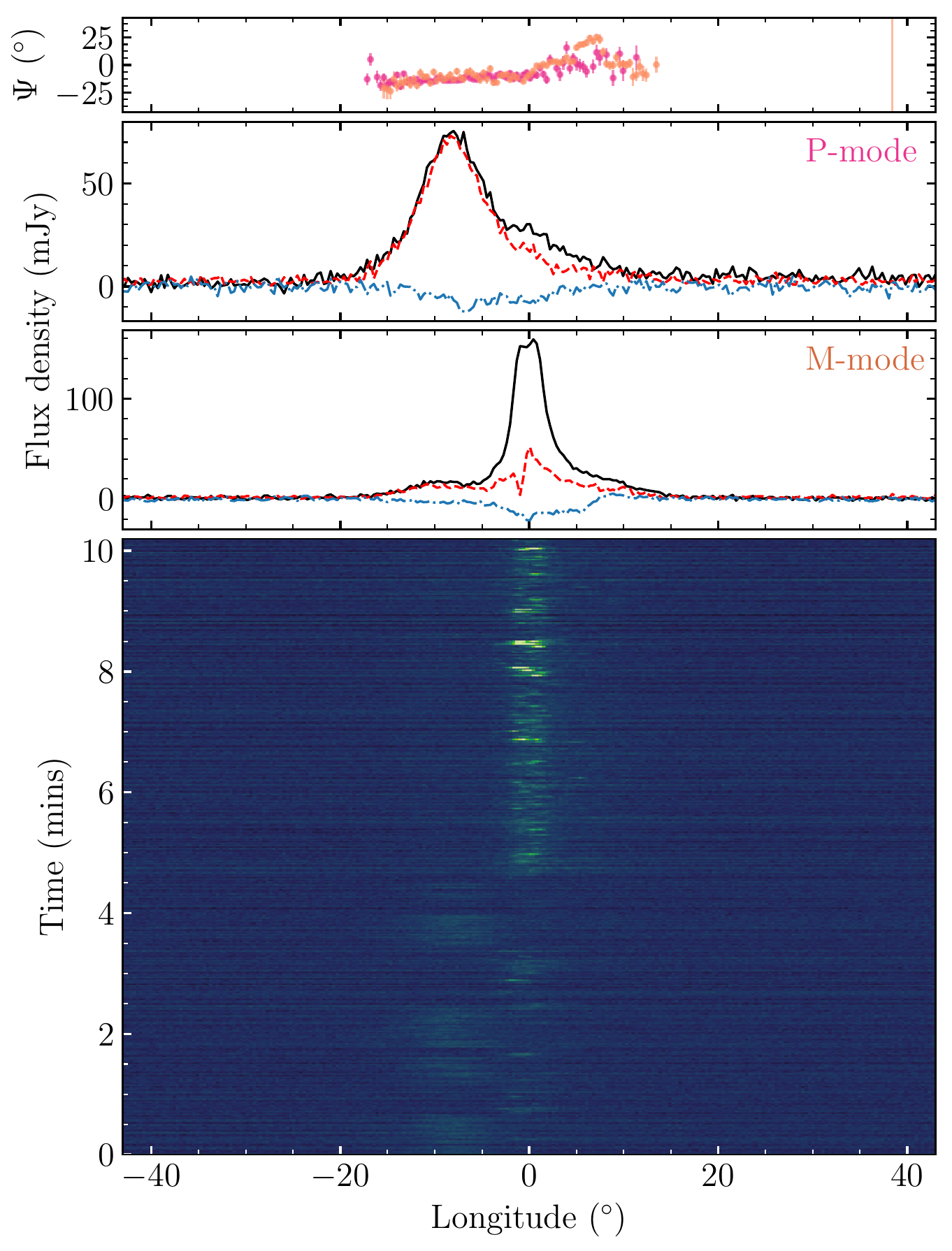}
    \caption{Same as Figure~\ref{fig:mode_switch_1}, but for the two emission modes observed on MJD 59047.}
    \label{fig:mode_switch_2}
\end{figure}

Inspecting the time-phase plot for MJD 59009 in Figure~\ref{fig:mode_switch_1}, it is clear the single-pulse emission of Swift J1818.0$-$1607 was quasi-periodically switching between a bright mode (B-mode) or a fainter quiet mode (Q-mode).
This is distinct from the largely random variations in single pulse flux and jitter often detected in magnetar single pulses \citep[e.g.][]{Serylak2009, Maan2019}.
Comparing the polarization profiles in Figure~\ref{fig:mode_switch_1}, the B-mode resembles the single component profile presented in \citet{Lower2020}, while the Q-mode is comprised of marginally detected emission at the same longitude as the B-mode and a slightly depolarized bump situated at approximately $+10^{\circ}$.
This bump in the Q-mode profile is positioned at the same pulse longitude as the peak of the inverted-spectrum component detected in later observations and close to the longitude of a depolarized bump seen at high frequencies in \citet{Lower2020} that was previously dismissed as an artefact from residual impulsive RFI. 
We also detected 10 pulses from a highly intermittent secondary profile component.
The fourth panel of Figure~\ref{fig:mode_switch_1} shows the resulting polarization profile after averaging together the single-pulse archives containing these bright two-component pulses.
Both the delay in pulse phase from the primary component and $\sim$60\,degree offset in the PA swing were seen in a secondary profile component detected by \citet{Champion2020} around the time of a glitch-like timing event in March 2020, possibly pointing to an increased level of rotational instability around the time of this observation.
A plateau in the B-mode profile at the pulse phase the secondary component points to faint pulses from this profile component appearing throughout the observation.
In total, we observed 295 rotations spent in the Q-mode and 157 in the B-mode.

We also detected emission mode switching on MJD 59047, however instead of the previous switching between a B- and Q-mode, the time-phase plot shown in Figure~\ref{fig:mode_switch_2} shows the magnetar varying between two longitudinally distinct modes. 
We termed these modes the P- and M-modes, as the spectrum of the P-mode resembles the steep spectra often seen in many rotation-powered pulsars while the M-mode exhibits the characteristically flat or inverted spectrum of radio-loud magnetars. 
Switching between these two modes was also detected by \citet{Pearlman2020ATel} who observed Swift J1818.0$-$1607 with the Deep Space Network two days prior (on MJD 59045) to us.
One marked difference to the emission mode switching detected in rotation-powered pulsars, is the mode-changing in Swift J1818.0$-$1607 was only a temporary phenomenon, as none of our subsequent observations show evidence for discrete switching between modes.
Instead, the magnetar appeared to remain in a constant M-mode-like emission state, suggesting whatever mechanism was driving the magnetospheric current variations had stabilised over the course of 15\,days.

\begin{figure}
    \centering 
    \includegraphics[width=\linewidth]{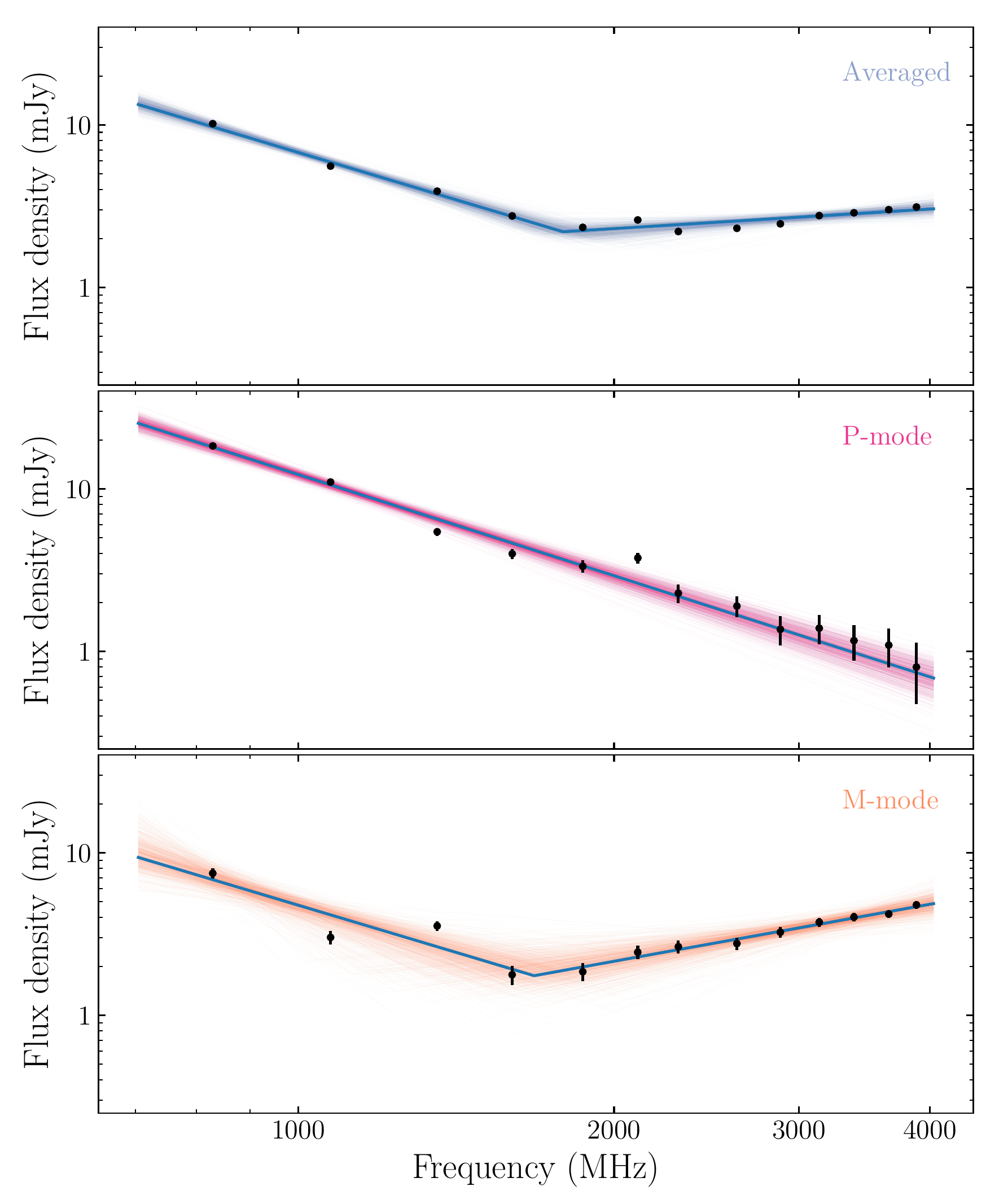}
    \caption{Continuum flux densities for the average of the two emission states (top), P-mode only (middle) and M-mode only (bottom). The solid-blue lines indicate the maximum likelihood posterior fit, while the coloured traces represent 1000 random draws from the posterior distributions.}
    \label{fig:mode_spectra}
\end{figure}

Using the spectrum fitting techniques outlined in Section~\ref{sec:spectra}, we found both emission modes detected on MJD 59009 are well described by a simple power-law with steep spectral indices of $\kappa = -2.6 \pm 0.1$ (B-mode) and $-4.5^{+1.2}_{-1.3}$ (Q-mode).
Note, the steeper spectrum of the Q-mode could be an artefact of the low S/N of this emission mode.
This is in contrast to the P- and M-modes detected on MJD 59047, where our flux density measurements and spectral fits shown in Figure~\ref{fig:mode_spectra} clearly show the P-mode has a steep spectrum with $\kappa = -2.0 \pm 0.1$ and the M-mode shows evidence of a spectral break at $\sim 1652$\,MHz and transition to an inverted spectrum, with corresponding spectral indices of $\kappa_{1} = -2.0^{+0.4}_{-0.6}$ and $\kappa_{2} = 1.2 \pm 0.3$. before and after the break. 
The consistency between the P-mode spectral index and M-mode pre-spectral break index indicates the magnetar continues to emit weak radio pulses from the P-mode-component while the M-mode is dominant.

Lastly, we checked for differences in the RM between the various emission modes that can arise from a variety of physical processes, such as the superposition of OPMs with different spectral indices and propagation effects within the neutron star magnetosphere \citep{Noutsos2009, Ilie2019}.
While we do successfully recover a RM of $ = 1441.0 \pm 0.6$\,rad\,m$^{-2}$ for the B-mode detected on MJD 59009, the RM was unconstrained for the Q-mode owing to the low level of emission associated with this mode.
For the P- and M-modes detected on MJD 59047, we obtained respective RM values of $1441.6 \pm 0.7$\,rad\,m$^{-2}$ and $1440.2 \pm 0.9$\,rad\,m$^{-2}$.
There is a significant amount of overlap between the posteriors for these two modes at the 68\,per cent confidence interval, suggesting any propagation effects between the the two modes are negligible.

%%%%%%%%%%%%%%%%%%%%%%%%%%%%%%%%%%%%%%%%%%%%%%%%%%
%%%%%%%%%%%%%%%%%%%%%%%%%%%%%%%%%%%%%%%%%%%%%%%%%%
\section{Polarization properties and geometry}\label{sec:pa_swing}

Earlier works noted Swift J1818.0$-$1607 possessed a relatively flat PA, potentially pointing to our line-of-sight only grazing the emission cone edge \citep{Lower2020, Champion2020}.
Similarly flat PAs were also detected across the first three observations shown in Figure~\ref{fig:obs}, however the PA swings across the last five epochs each differ dramatically, bearing a striking resemblance to the S-shaped swing expected from the simple rotating vector model (RVM) of \citet{Radhakrishnan1969}.
Under the RVM, the sweep of the PA is a purely geometric effect caused by the changing angle between the projected dipole magnetic-field direction and our line of sight. 
It can be expressed in terms of the magnetic inclination angle ($\alpha$; the angle between the spin and magnetic axes) and the angle between the spin axis and our line of sight ($\zeta$) as
\begin{equation}\label{eqn:rvm}
    \tan(\Psi - \Psi_{0}) = \frac{\sin\alpha \sin(\phi - \phi_{0})}{\sin\zeta \cos\alpha - \cos\zeta \sin\alpha \cos(\phi - \phi_{0})},
\end{equation}
where $\phi_{0}$ is the pulse longitude at which $\Psi = \Psi_{0}$, i.e, the PA of the pulsar spin axis projected onto the plane of the sky.
The difference between $\zeta$ and $\alpha$ is the angle of closest approach between our line of sight and the magnetic axis ($\beta = \zeta - \alpha$), hereafter referred to as the magnetic impact angle.
While the RVM is only truly valid in the case of an unchanging, axisymmetric dipole magnetic field, the geometric interpretation of the model can potentially provide some insight to the processes driving the PA variations \citep{Everett2001, Johnston2019b}.

\begin{figure}
    \centering    
    \includegraphics[width=\linewidth]{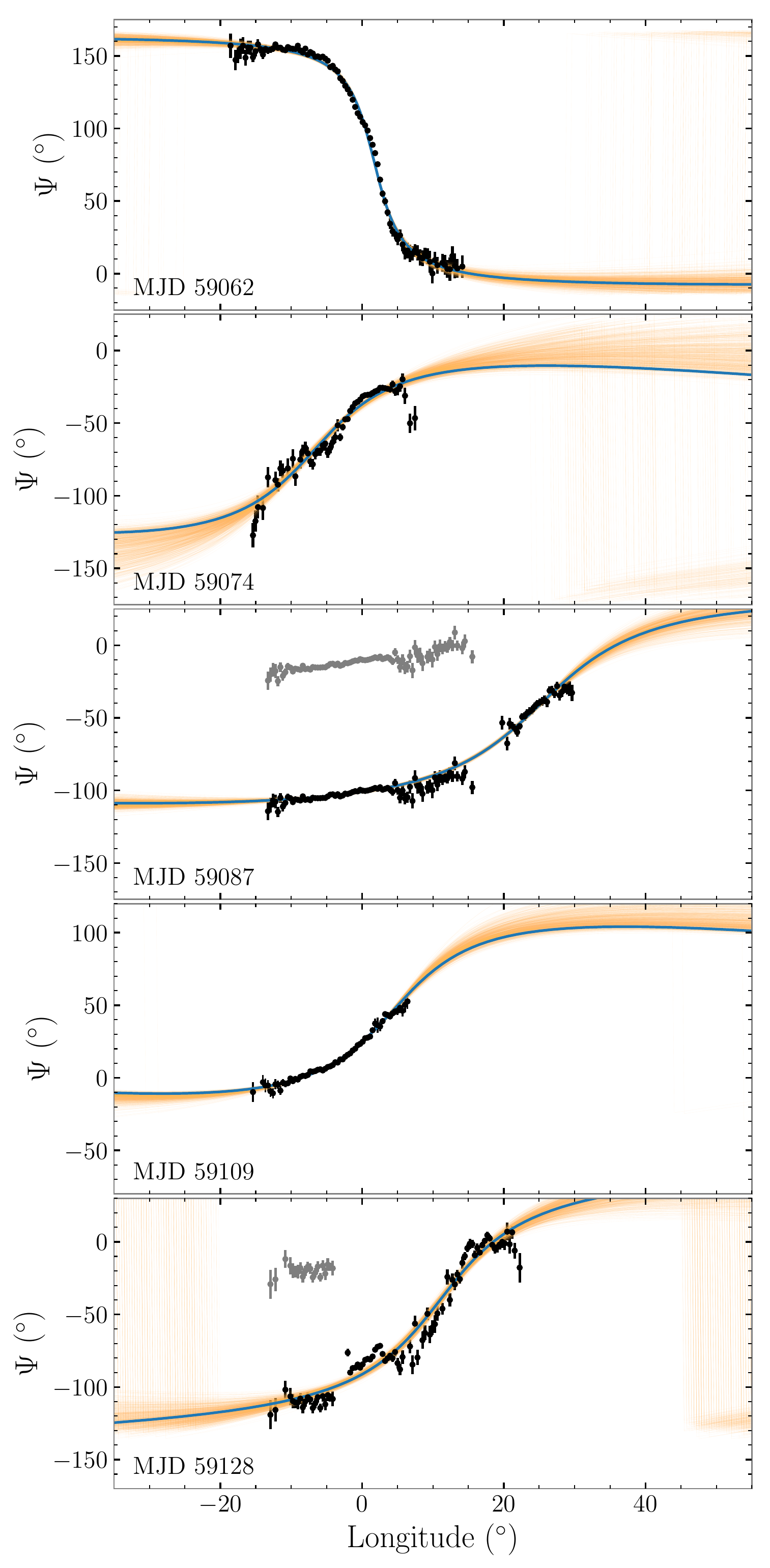}
    \caption{RVM fits to the five RVM-like PA swings (black points) with the maximum likelihood a posteriori fit (blue) along with traces generated from 1000 random draws from the posterior distributions (orange). Grey points indicate PA values prior to adding an OPM correction.}
    \label{fig:RVM_fits}
\end{figure}

\begin{figure}
    \centering
    \begin{subfigure}[b]{1\linewidth}
        \centering
        \includegraphics[width=\linewidth]{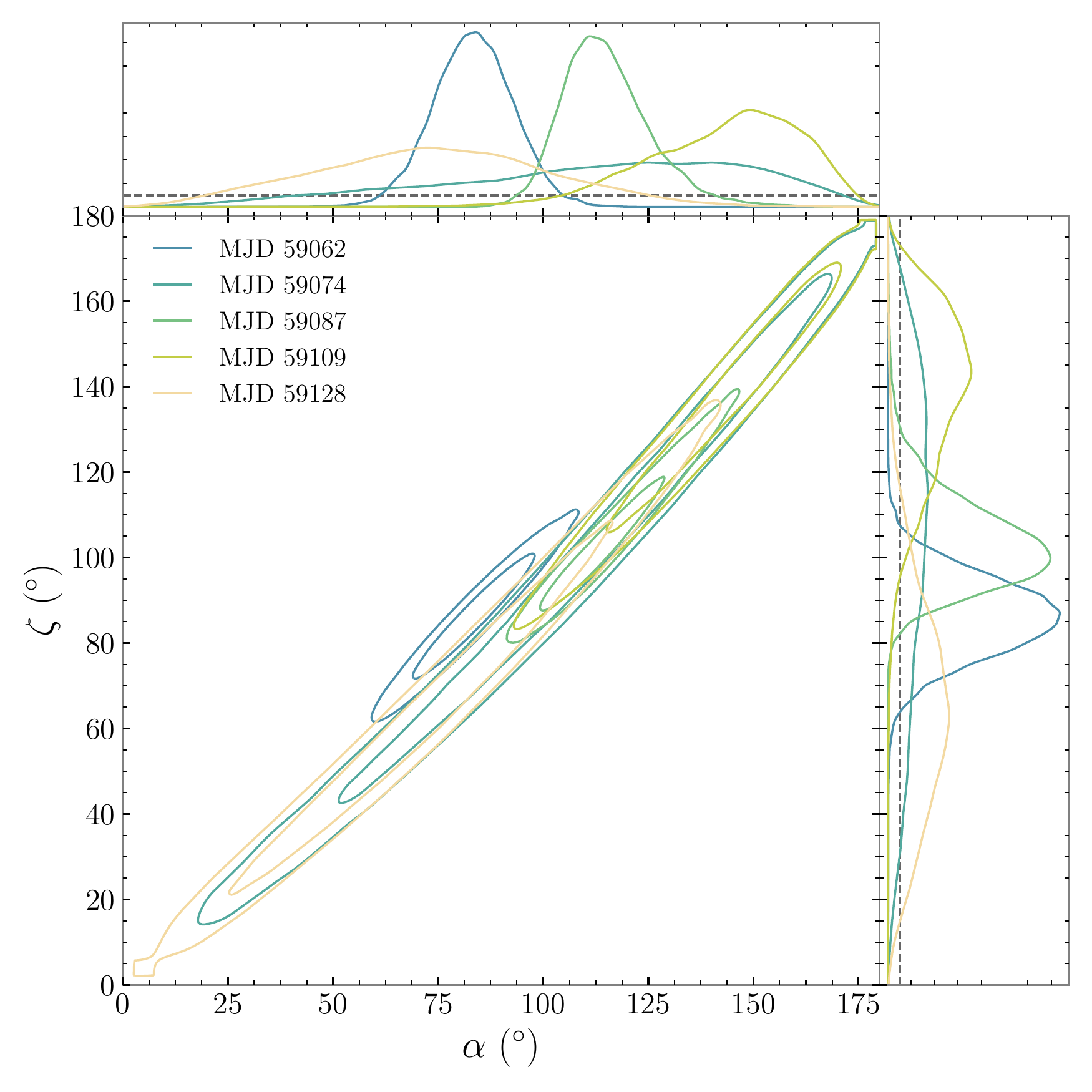}
    \end{subfigure}
    \begin{subfigure}[b]{1\linewidth}
        \centering
        \includegraphics[width=\linewidth]{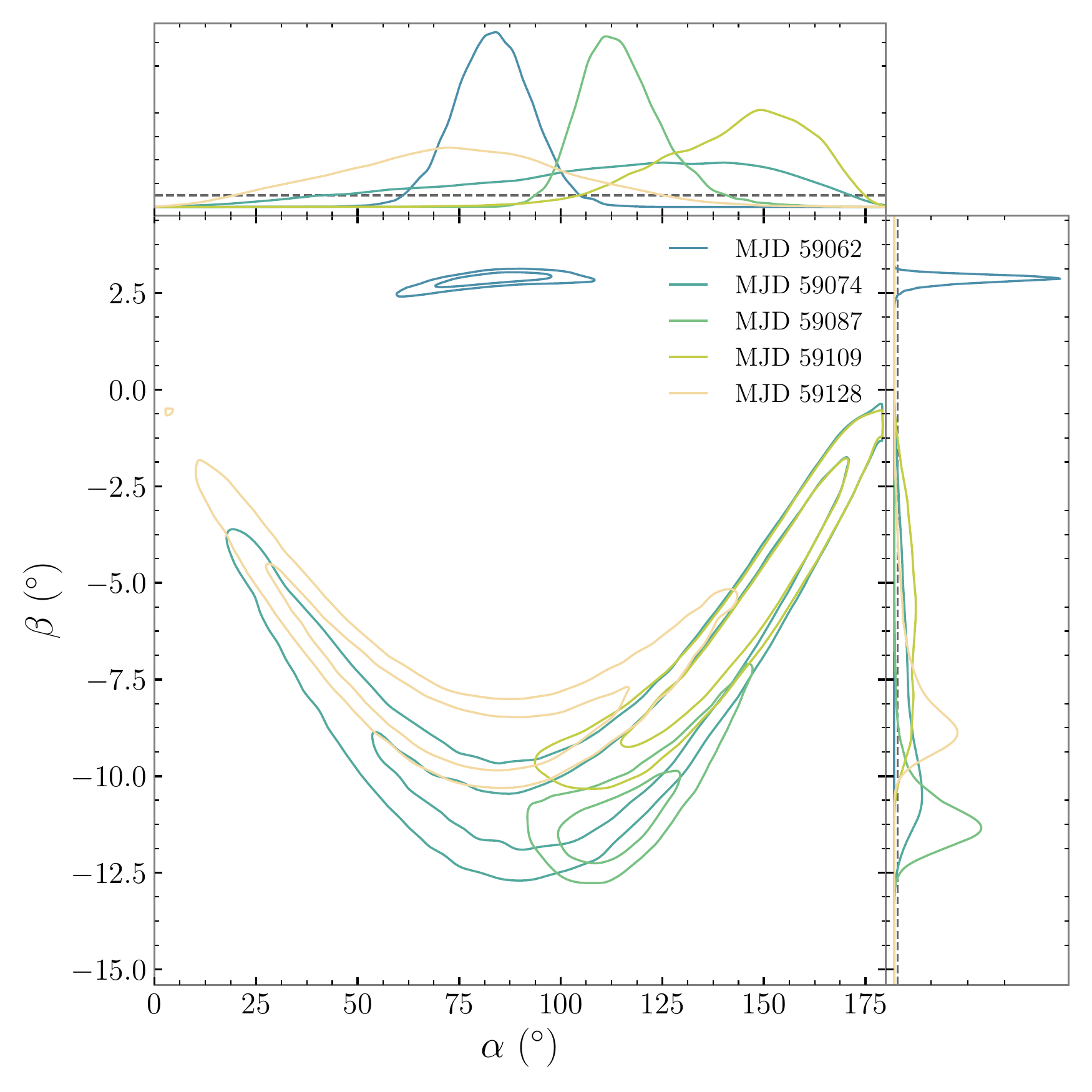}
    \end{subfigure}
    \caption{Comparison between the one- and two-dimensional posterior distributions of $\alpha$ and $\zeta$ (top) and $\alpha$ and $\beta$ (bottom). Contours indicate the 68\% and 95\% confidence regions. Grey-dashed lines in the one-dimensional posteriors indicate the priors.}
    \label{fig:rvm_post}
\end{figure}

%%%%%%%%%%%%%%%%%%%%%%%%%%%%%%%%%%%%%%%%%%%%%%%%%%
\subsection{Viewing geometry}
\begin{table}
\begin{center}
\caption{RVM-fits to the data. MJDs with a $^{\dagger}$ include corrections for orthogonal polarization modes.\label{tbl:rvm}}
\renewcommand{\arraystretch}{1.2}
\setlength{\tabcolsep}{4pt}
\resizebox{\linewidth}{!}{
\begin{tabular}{lccccc}
\hline
MJD & $\alpha$     & $\beta$      & $\phi_{0}$   & $\Psi_{0}$   & $\zeta$      \\
    & ($^{\circ}$) & ($^{\circ}$) & ($^{\circ}$) & ($^{\circ}$) & ($^{\circ}$) \\
\hline
58977 & $106^{+29}_{-36}$ & $-67^{+30}_{-41}$ & $-3^{+35}_{-32}$ & $15^{+37}_{-35}$ & $24^{+44}_{-18}$ \\
59009 & $93^{+39}_{-31}$ & $-37^{+14}_{-20}$ & $20^{+14}_{-19}$ & $25^{+21}_{-26}$ & $49^{+58}_{-36}$ \\
59047 & $106^{+23}_{-29}$ & $-71^{+30}_{-32}$ & $4^{+33}_{-42}$ & $2^{+36}_{-42}$ & $20^{+45}_{-16}$ \\
59062 & $82 \pm 9.4$ & $2.80^{+0.08}_{-0.13}$ & $1.69 \pm 0.08$ & $77.2^{+0.9}_{-0.8}$ & $85 \pm 10$ \\
59074 & $115^{+31}_{-46}$ & $-9^{+4}_{-2}$ & $-7 \pm 1$ & $-71^{+4}_{-6}$ & $104^{+37}_{-46}$ \\
59087$^{\dagger}$ & $113^{+11}_{-9}$ & $-11.2^{+0.9}_{-0.6}$ & $25.3^{+0.7}_{-0.8}$ & $-40 \pm 3$ & $102^{+12}_{-9}$ \\
59109 & $144^{+14}_{-22}$ & $-6 \pm 3$ & $4.7^{+0.5}_{-0.3}$ & $-48^{+3}_{-2}$ & $138^{+19}_{-25}$ \\
59128$^{\dagger}$ & $71 \pm 28$ & $-8.5^{+2.0}_{-0.8}$ & $11.1 \pm 0.8$ & $-40^{+5}_{-3}$ & $62^{+28}_{-26}$ \\
\hline
\end{tabular}
}
\renewcommand{\arraystretch}{}
\end{center}
\end{table}

Using the RVM and a Gaussian likelihood function, we fitted each of the PA swings shown in Figure~\ref{fig:obs}. 
We assumed uniform priors on all RVM parameters, except for $\phi_{0}$ where we employed a Gaussian prior centred at $0^{\circ}$ with a width of $45^{\circ}$.
This constrained prior allows us to avoid the ambiguity in which magnetic pole the polarized radio emission originates, as we do not know the sense of the magnetar's rotation.
For the observations on MJD 59062 and 59074 we applied a $+180^{\circ}$ and $-180^{\circ}$ phase jump respectively to PA values below $-6^{\circ}$ in order to have a smooth PA swing across the pulse profile. 
We also corrected the $90^{\circ}$ jump in the PA swings on MJD 59087 and 59128 due to OPM transitions by subtracting $-90^{\circ}$ from the measured PA values at $\phi \leq 9^{\circ}$.
The results of our RVM-fits are presented in Table~\ref{tbl:rvm}. 
In Figure~\ref{fig:RVM_fits} we show the PA swings from the last five epochs; our overlaid RVM-fits are in excellent agreement with the data. 

Our best constrained values of the geometry from the MJD 59062 ($\alpha = 82^{\circ}$, $\beta=3^{\circ}$) and 59087 ($\alpha = 112^{\circ}$, $\beta=-11^{\circ}$) observations are highly inconsistent, as could already be discerned from the opposite sweep of the PAs on these dates.
The relatively flat PA swings and narrow pulse duty-cycles seen on MJD 58977, 59009 and 59047 resulted in our recovered values for $\alpha$ and $\zeta$ being relatively unconstrained although the positive PA gradients indicate $\beta<0$ at these epochs.
Figure~\ref{fig:rvm_post} shows the one- and two-dimensional posterior distributions of $\alpha$, $\zeta$ and $\beta$ from our fits to the PAs in Figure~\ref{fig:RVM_fits}. 
It is clear that both $\alpha$ and $\zeta$ remain largely consistent between the four observations that show positive PA gradients.
As the marginalised $\zeta$ posteriors for MJD 59062 and 59087 share a significant amount of overlap at the 68\% confidence level and the $\alpha$ posteriors do not, the most likely explanation for the flipped PA swing and inferred $\beta$ on MJD 59062 is a sudden change in $\alpha$ that occurred between MJD 59047 and MJD 59062, that subsequently reversed sometime prior to MJD 59074.
We discuss the implications and describe probable causes of this effect below in Section~\ref{sec:disc}. 
If we ignore the results from MJD 59062, then we can combine the $\alpha$ and $\zeta$ posteriors at every other epoch to obtain improved measurements of $\alpha = 112^{\circ}\,^{+7}_{-9}$ and $\zeta = 99^{\circ}\,^{+7}_{-10}$, which in turn correspond to $\beta = -12.9^{\circ}\,^{+0.6}_{-0.7}$.

Given the inferred geometry, the radio pulses must originate from high above the neutron star surface in order to explain the average profile width (at the 10\% flux level) of $W_{10, {\rm avg}} = 34.1^{\circ}$.
Using the measured values of $\alpha$ and $\zeta$, we can infer a minimum geometric emission height by first computing the half-opening angle of the emission cone ($\rho$; \citealt{Gil1984})
\begin{equation}\label{eqn:h_em_geo_1}
    \cos\rho = \cos\alpha \cos\zeta + \sin\alpha \sin\zeta \cos(W/2),
\end{equation}
where $W$ is the pulse width, taken to be $W_{10}$ (in units of rad) in our case.
Assuming the emission extends to the last open field line and a fully active polar cap with symmetric emission about the pole, the emission height, $h_{\rm em}$, can be derived via \citep{Rankin1990}
\begin{equation}\label{eqn:h_em_geo_2}
    \rho = 3 \sqrt{\frac{\pi h_{\rm em}}{2 P c}},
\end{equation}
where $P = 1.3635$\,s is the spin period of the magnetar and $c$ is the vacuum speed of light.
From the inferred geometry and $W_{10}$ we obtain a minimum emission height of $3800$\,km, i.e close to $6$\,per cent the light-cylinder radius of Swift J1818.0$-$1607 ($r_{\rm lc} = 6.5 \times 10^{4}$\,km). 

\begin{figure}
    \centering    
    \includegraphics[width=\linewidth]{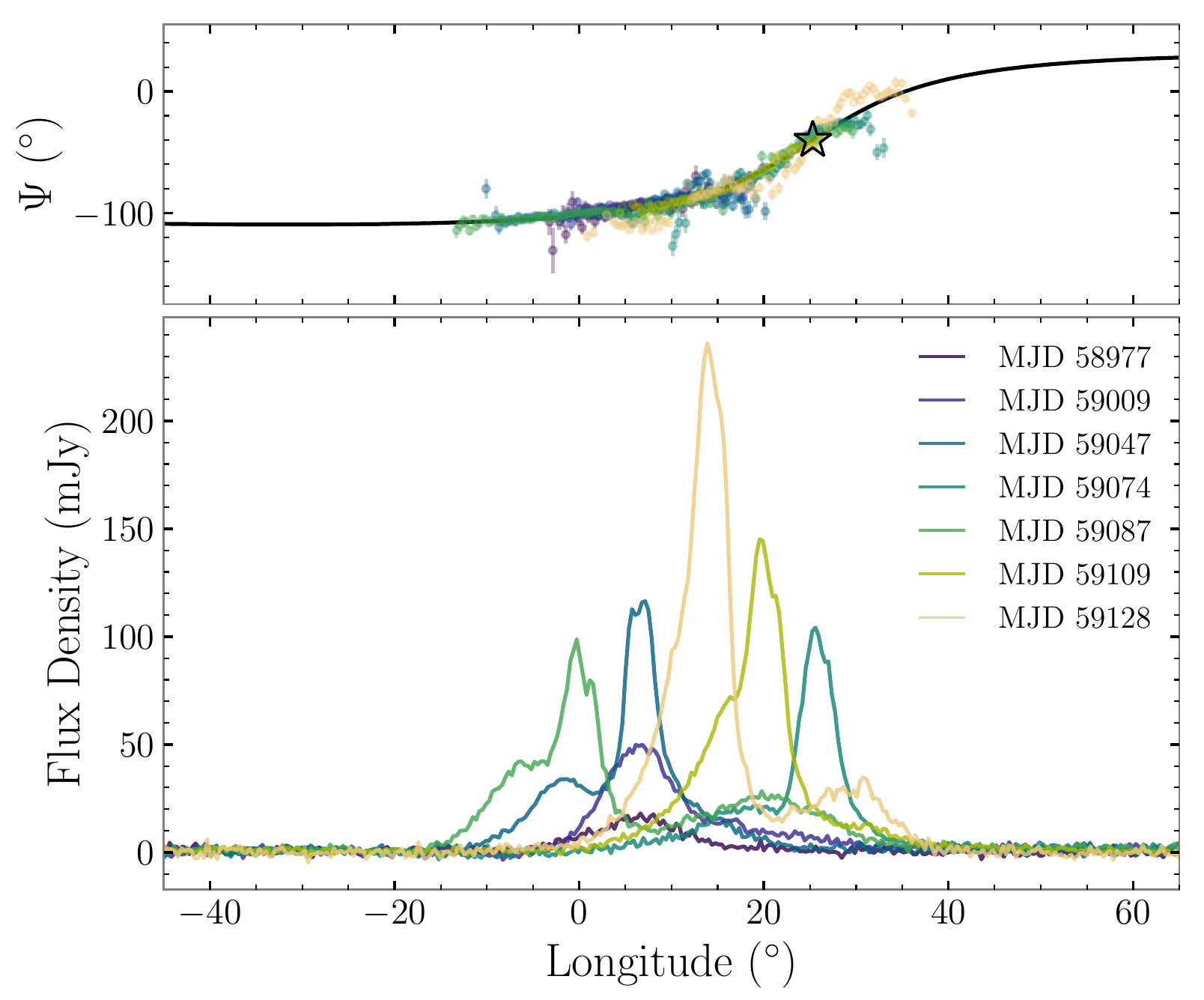}
    \caption{Comparison of the PA swings (top) and total intensity profiles (bottom) of the PA aligned profiles. The black coloured line and star in the top panel represents the median RVM-fit and corresponding position of ($\phi_{0}$, $\Psi_{0}$) for the PA swing on MJD 59087.}
    \label{fig:overlap}
\end{figure}

%%%%%%%%%%%%%%%%%%%%%%%%%%%%%%%%%%%%%%%%%%%%%%%%%%
\subsection{Position angle alignment and emission heights}

Visually, the PA swings that are shown for MJD 59074 onward in Figure~\ref{fig:RVM_fits} appear similar, and could easily be aligned by the addition of $90^{\circ}$ jumps in PA and shifts in pulse longitude. A similar argument could be made for all of the PA swings prior to MJD 59062 since they all show evidence of shallow positive PA gradients.
To investigate whether it is possible to align the PAs , we first visually aligned each PA swing by adding a $-135^{\circ}$ jump to MJD 58977 and $-90^{\circ}$ jumps to MJD 59009, 59047 and 59109 respectively.
We then measured the longitude offsets required to align the PAs by performing a least-squares fit to the data assuming values of $(\alpha, \zeta) = (112^{\circ}, 99^{\circ})$ from the combined posteriors, and ($\phi_{0}, \Psi_{0}) = (25.5^{\circ}, -40.2^{\circ})$ from the RVM fit to MJD 59087.
The PAs and total intensity profiles after applying the resulting phase offsets are displayed in Figure~\ref{fig:overlap}.
There are two possible ways to interpret the PA aligned profiles: longitudinal wandering or oscillating of the emission patch over time, or temporal evolution of the emission height. 
Longitudinal motion of the emission patch would imply the the magnetar possesses a largely unfilled emission cone with an opening angle that is difficult to reconcile with our inferred magnetic geometry and its rotation period of $\sim$1.4\,s.
Hence, for the remainder of this section we focus on the more likely possibility of a changing emission height.

Pulsar emission theory predicts radiation produced nearer to the magnetic pole will originate from magnetic field lines closer to the neutron star surface \citep[see, e.g.][]{Yuen2014}.
If the emission region is symmetric about the magnetic meridian, then relativistic aberration and retardation effects will cause the observed PA to lag behind the total intensity profile \citep{Blaskiewicz1991}.
The emission height relative to the centre of the neutron star ($h_{\rm em}$) can be inferred from the magnitude of this delay expressed in terms of pulse longitude ($\delta\phi$, in units of rad) and radius of the light cylinder, $r_{\rm lc}$, as
\begin{equation}\label{eqn:h_em}
    h_{\rm em} = \frac{r_{\rm lc}}{4}\delta\phi = \frac{P\,c}{8\pi}\delta\phi.
\end{equation}

Figure~\ref{fig:overlap} shows that the observation on MJD 59074 must have the lowest emission height, as the pulse profile is almost aligned with the inflexion point of the RVM fit. 
We therefore take this observation as a reference for computing relative emission heights noting that the absolute height is difficult to ascertain as the location on the profile of the pole crossing is unclear.
Table~\ref{tbl:h_em} shows the longitude offset between the profiles relative to the observation on MJD 59074 and hence the inferred relative values of $h_{\rm em}$ expressed in both km and as a fraction of the light cylinder radius. 
We choose not to include uncertainties because the main point is to demonstrate indicative changes in emission height.
The table shows there is substantial variation in emission height between the epochs and no particular trend with time. 
Why the emission height should change in this way is unclear, but the fact that the profile components persist implies that the same field lines are being illuminated over the range of heights.

\begin{table}
\begin{center}
\caption{Longitude offsets and relative emission heights.\label{tbl:h_em}}
\renewcommand{\arraystretch}{1.2}
\setlength{\tabcolsep}{4pt}
\begin{tabular}{lccc}
\hline
MJD & $\delta\phi$ & $h_{\rm em} - h_{\rm em, MJD\,59074}$ & $(h_{\rm em} - h_{\rm em, MJD\,59074})/r_{\rm lc}$ \\
    & ($^{\circ}$) & (km)                                  &                                                             \\
\hline
58977 & $18.6$ & $5200$  & $0.08$ \\
59009 & $19.1$ & $5400$  & $0.08$ \\
59047 & $18.8$ & $5300$  & $0.08$ \\
59074 & $0$    & $0$     & $0$    \\
59087 & $25.5$ & $7200$  & $0.11$ \\
59109 & $5.7$  & $1600$  & $0.02$ \\
59128 & $11.7$ & $3300$  & $0.05$ \\
\hline
\end{tabular}
\renewcommand{\arraystretch}{}
\end{center}
\end{table}

It is evident the polarization profiles corresponding to epochs with smaller inferred emission heights possess the largest variations in polarization fraction.
For instance, the leading, steep-spectrum component normally has close to 100\,percent linear polarization, however on MJD 59074 -- the observation with the smallest inferred emission height -- its polarization fraction is less than half of what it is at every other epoch. 
This is consistent with observations of rotation-powered pulsars, where radio emission emitted lower in the magnetosphere is more likely to be subject to a larger amount of magnetospheric propagation effects \citep[e.g.][]{Smith2013}.

\begin{figure}
    \centering    
    \includegraphics[width=\linewidth]{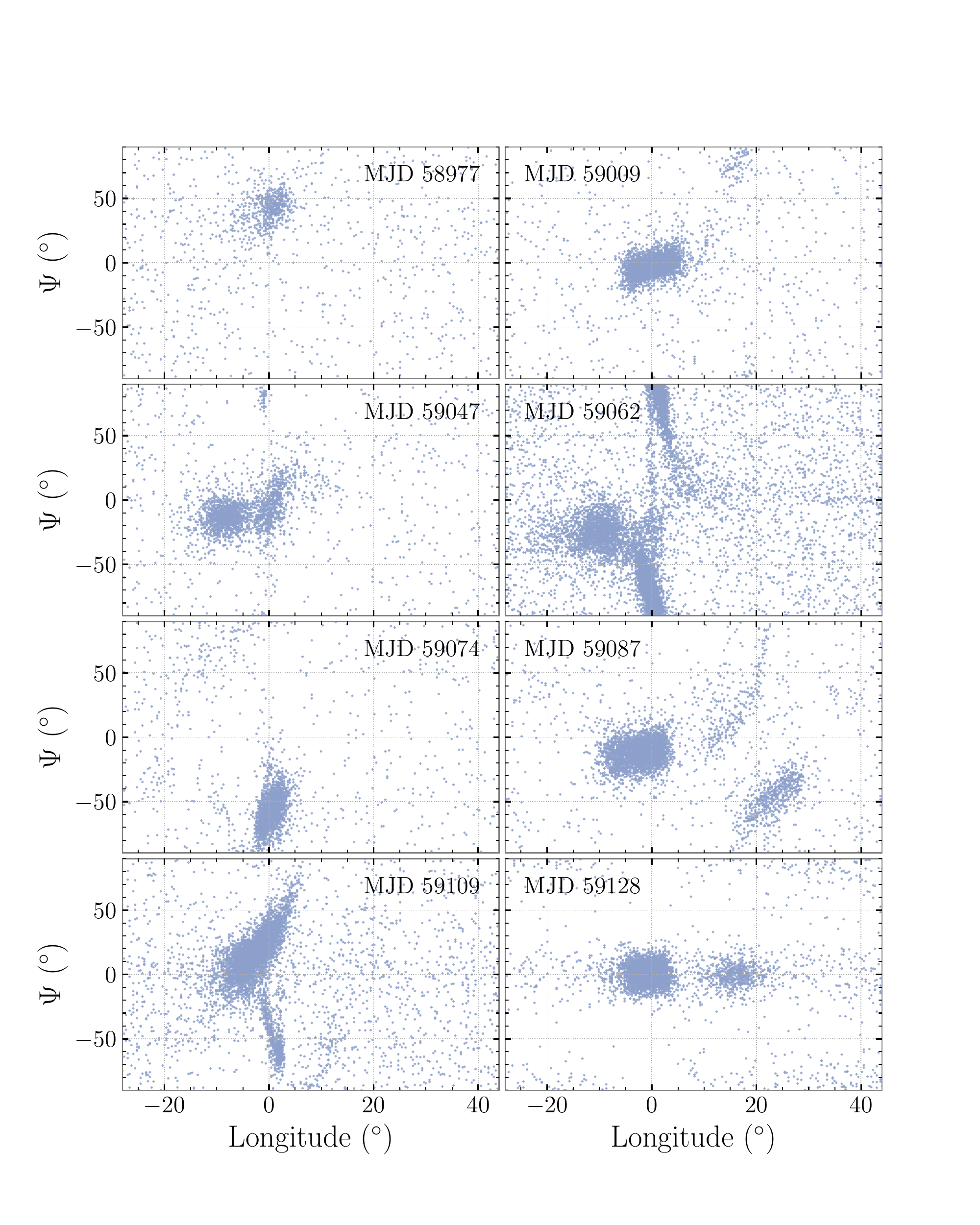}
    \caption{Pulse and phase resolved polarization position angle ($\Psi$).}
    \label{fig:poln_modes}
\end{figure}

%%%%%%%%%%%%%%%%%%%%%%%%%%%%%%%%%%%%%%%%%%%%%%%%%%
%%%%%%%%%%%%%%%%%%%%%%%%%%%%%%%%%%%%%%%%%%%%%%%%%%
\subsection{Polarization modes}

Many normal pulsars emit linearly polarized radio waves in two modes that are usually orthogonal to one another \citep{Backer1976}.
It is believed these modes arise from propagation effects within the neutron star magnetosphere, such as refraction and birefringence \citep{McKinnon1997, Petrova2001}.
If two or more OPMs exist, then the process of averaging over many rotations can suppress the observed linear polarization in pulsar profiles \citep[e.g.][]{Karastergiou2002}.
This could explain the apparent depolarization of the inverted spectrum profile component detected on MJD 59047 onward.
We tested this idea by studying the distribution of PA values at each phase bin across the pulse profile.
To minimise spurious contributions from noise and unaccounted RFI, we imposed a threshold where the linear polarization of a given phase bin must be a factor of 2.5 times greater than the off-pulse RMS when calculating the PA.
Scatter plots showing the PA distributions for all eight epochs are presented in Figure~\ref{fig:poln_modes}. 

In general, the scatter plots largely follow the PA swings depicted in Figure~\ref{fig:obs}.
This is not surprising for the profiles/profile-components that show a high amount of linear polarization, as the presence of OPMs would result in depolarization.
Aside from the known OPM on MJD 59087, we find evidence of additional OPMs on MJD 59047 at longitudes between $-5^{\circ}$ to $0^{\circ}$, as indicated by the small cluster of points that have an approximately $+90^{\circ}$ offset in PA from the majority of the scatter plot, and a possible OPM on around longitudes close to $0^{\circ}$ on MJD 59062.
The offset clump of PA values at longitudes between $10$ to $20^{\circ}$ on MJD 59009 can be attributed to the handful of bright pulses detected from the secondary profile component shown in the fourth panel of Figure~\ref{fig:mode_switch_1}. 
In general, there is a notable lack of additional polarization modes in the components that have low linear polarization fractions.
This could be due to pulses displaying emission from additional modes being intermittent and we simply did not catch a large amount of these pulses in our relatively short observations.
Longer observations performed by other facilities may be able to place stronger constraints on the presence of any additional polarization modes.

While the majority of the PA distribution on MJD 59109 follows the expected curve seen in the average PA, there are a number of points that follow a branching PA swing that bends away from the bulk distribution.
Intriguingly, the slope of the branch appears to match that of the PA swing (and overall PA distribution) observed on MJD 59062. 
Similar branching behaviour has been seen in the PA distributions of some rotation-powered pulsars \citep[e.g. Figure 4 of][]{Ilie2020}.
Remarkably, after visually aligning of the two PA distributions by adding a $+3^{\circ}$ offset to MJD 59062, the slope of the drifting branch matches the downward portion of the PA distribution of MJD 59062. 
This naturally raises the question: did we observe sporadic pulses from the same, reversed PA emission mode that was detected on MJD 59062?

\begin{figure}
    \centering    
    \includegraphics[width=\linewidth]{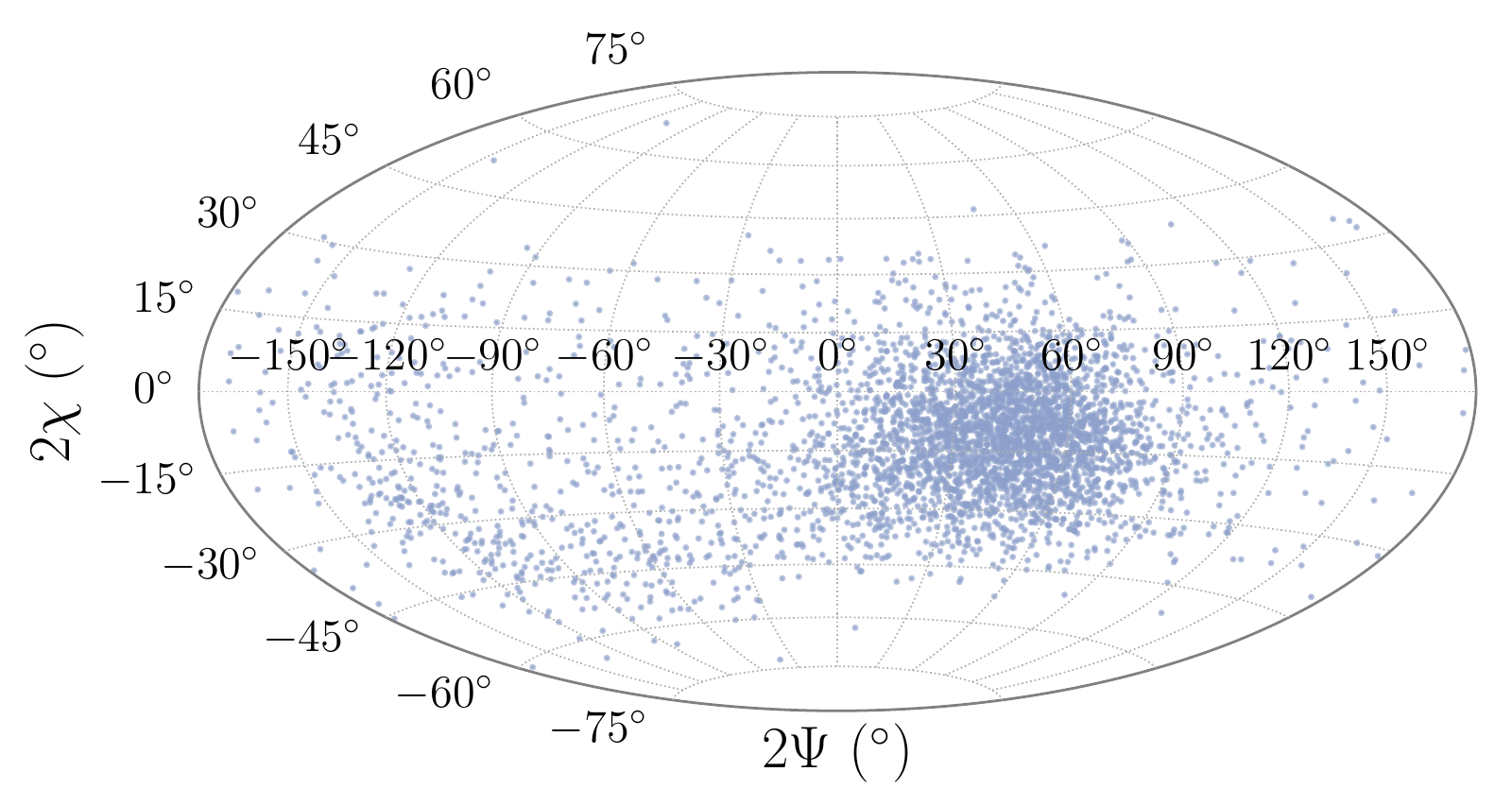}
    \caption{Hammer equal-area projection of the polarization position angle ($\Psi$) and ellipticity angle ($\chi$) distributions from MJD 59109 on the Poincar{\'e} sphere.}
    \label{fig:200917_hammer}
\end{figure}

Inspection of individual pulses associated with this `abnormal' polarization mode reveals the majority exhibit PAs with a continuous downward drift as a function of pulse longitude, while a handful show evidence of an initially upward drifting PA followed by an apparent OPM jump to the tail of the downward-drifting PA distribution. 
In general, they all show a lower linear polarization fraction compared to pulses from the `normal' mode and significantly increased amounts of circular polarization.
\citet{Dyks2020} devised a phenomenological model in which similar behaviour can originate from the passage of the emission patch along a great circle close to one of the Stokes $V$ poles when projected onto the Poincar{\'e} sphere. 
We tested whether such a passage is present within our data by visually inspecting the position angle and corresponding ellipticity angle distributions plotted on the Poincar{\'e} sphere in Figure~\ref{fig:200917_hammer}.
Most of the polarization distribution is concentrated in a blob centred near $(2\Psi, 2\chi) = (45^{\circ}, -15^{\circ})$, however the low-density distribution with negative values of $\Psi$, i.e values associated with the downward drifting branch, appears to trace out a rough circular pattern similar to those presented in Figures 2 and 3 of \citet{Dyks2020}.
This suggests the pulses associated with the downward-drifting PA branch are not associated with the reversed PA swing detected on MJD 59062.
Instead they may represent a sample of pulses that experienced a propagation effect within the magnetosphere that masquerades as a smeared OPM in the 2-dimensional PA-longitude plot.

%%%%%%%%%%%%%%%%%%%%%%%%%%%%%%%%%%%%%%%%%%%%%%%%%%
%%%%%%%%%%%%%%%%%%%%%%%%%%%%%%%%%%%%%%%%%%%%%%%%%%
\section{Discussion}\label{sec:disc}

Radio-loud magnetars are unusual in that their flat or inverted spectra means they are detectable as pulsars at millimetre-wavelengths \citep[e.g.][]{Camilo2007b}.
Hence it was surprising when Swift J1818.0$-$1607 was found to possess a steep, negative spectral index.
However, given its similarities to the population of high B-field pulsars, we speculated in \citet{Lower2020} that the current radio outburst may progress in a similar fashion to the 2016 magnetar-like outburst of PSR J1119$-$6127 \citep{Majid2017}, and the spectrum could begin to flatten over the months following its discovery.
An earlier, multi-wavelength observation by the Deep Space Network (MJD 58947) reported a possible flattening of the spectrum \citep{Majid2020ATelA}, and the apparent trend toward smaller spectral indices over time found by \citet{Champion2020} seemingly pointed to the spectrum following this prediction.
However, our first two spectral index measurements listed in Table~\ref{tbl:spectral} appear to be in conflict with this hypothesis, and it was only after the emergence of a new profile component bearing an inverted spectrum that the phase-averaged spectral index showed any sign of flattening. 
Our measured spectral index for this component ranges between $\kappa = -0.2$ to $+0.7$, similar to the those of other radio-loud magnetars \citep{Levin2012, Dai2019}, and enabled pulses from Swift J1818.0$-$1607 to be detected up to millimetre wavelengths \citep{Torne2020ATel}.
Intriguingly the negative reported spectral index at these high wavelengths, combined with flatter, but still negative spectral indices measured between 6 and 39\,GHz by Effelsberg and the Deep Space Network \citep{Liu2020ATel, Pearlman2020ATel}, indicate Swift J1818.0$-$1607 possesses a high-frequency spectral turnover.
Both SGR 1745$-$2900 and XTE J1810$-$197 were detected at similarly high radio frequencies following their 2013 and 2003/2018 outbursts \citep{Torne2015, Pennucci2015, Camilo2007b, Torne2020} and also showed evidence of similar spectral behaviour, indicating high-frequency turnovers may be a common feature of the magnetar radio emission mechanism.

In addition to developing a flat-spectrum component, we also detected two distinct types of emission mode switching at two separate epochs, along with dramatic variations in the position angle swing.
However, Swift J1818.0$-$1607 is not the only magnetar found to exhibit mode switching.
The mode changes in the single pulses from SGR 1745$-$2900 are a subtle effect, manifesting as slight changes in the leading edge of its profile \citep[see Figure 3 of][]{Yan2018}.
In contrast, 1E~1547.0$-$5408 has been seen to undergo at least two types of transient profile events: bright bursts followed by emission appearing at slightly earlier pulse longitudes before `recovering' back to its initial position \citep[Figure 2 of][]{Camilo2007a}, and discrete switching to and from an emission mode where the profile grows an extra hump on its trailing shoulder \citep[see Figure 5 of][]{Halpern2008}.
The latter mode appears somewhat similar to the P- and M-mode switching we detected on MJD 59047, however the lack of spectral analyses of the 1E~1547.0$-$5408 precludes a more direct comparison.
We can however draw some parallels between the modes of Swift J1818.0$-$1607 and the curious behaviour of the high B-field pulsar PSR J1119$-$6127, where a number of one-off profile variations were observed by \citet{Weltevrede2011} following a large glitch in 2007.  
This included a transient secondary profile component that lags the primary by $\sim 30^{\circ}$, bearing a somewhat similar profile shape and polarization fraction to the secondary component we detected in Swift J1818.0$-$1607 on MJD 59009. 
They also detected highly sporadic pulses similar to those from rotating radio transients (RRATs) during two separate epochs where the pulsar was observed at two different frequency bands: 4 bright pulses during a 20-cm observation, and a handful at 10-cm.
It was argued the rate at 10-cm must be much higher than at 20-cm as the pulsar was rarely observed at this frequency band.
This apparent increased detection rate could be interpreted as the RRAT-like pulses possessing a more magnetar-like, inverted spectral index. 
Intriguingly, no reported enhancement to the pulsars X-ray emission was associated with this glitch \citep{Gogus2016}, unlike the 2016 glitch that was associated with a magnetar-like outburst \citep{Archibald2016}.
\citet{Dai2018} found PSR J1119$-$6127 exhibited dramatic variations in its polarisation properties during the 2016 outburst, in particular the transient secondary component, showed a similar amount of polarization variations as Swift J1818.0$-$1607.
One notable difference between the polarization variations in Swift J1818.0$-$1607 and PSR J1119$-$6127 is the latter showed extreme deviations from its normally flat PA swing, exhibiting a variety of non-RVM-like variations over the course of a few days. 
Similar strong variations in the polarization fraction and PA swing of XTE J1810$-$197 were observed after its 2018 outburst that again deviate significantly from the predictions of a simple RVM model \citep{Dai2019}.

\begin{figure}
    \centering    
    \includegraphics[width=0.9\linewidth]{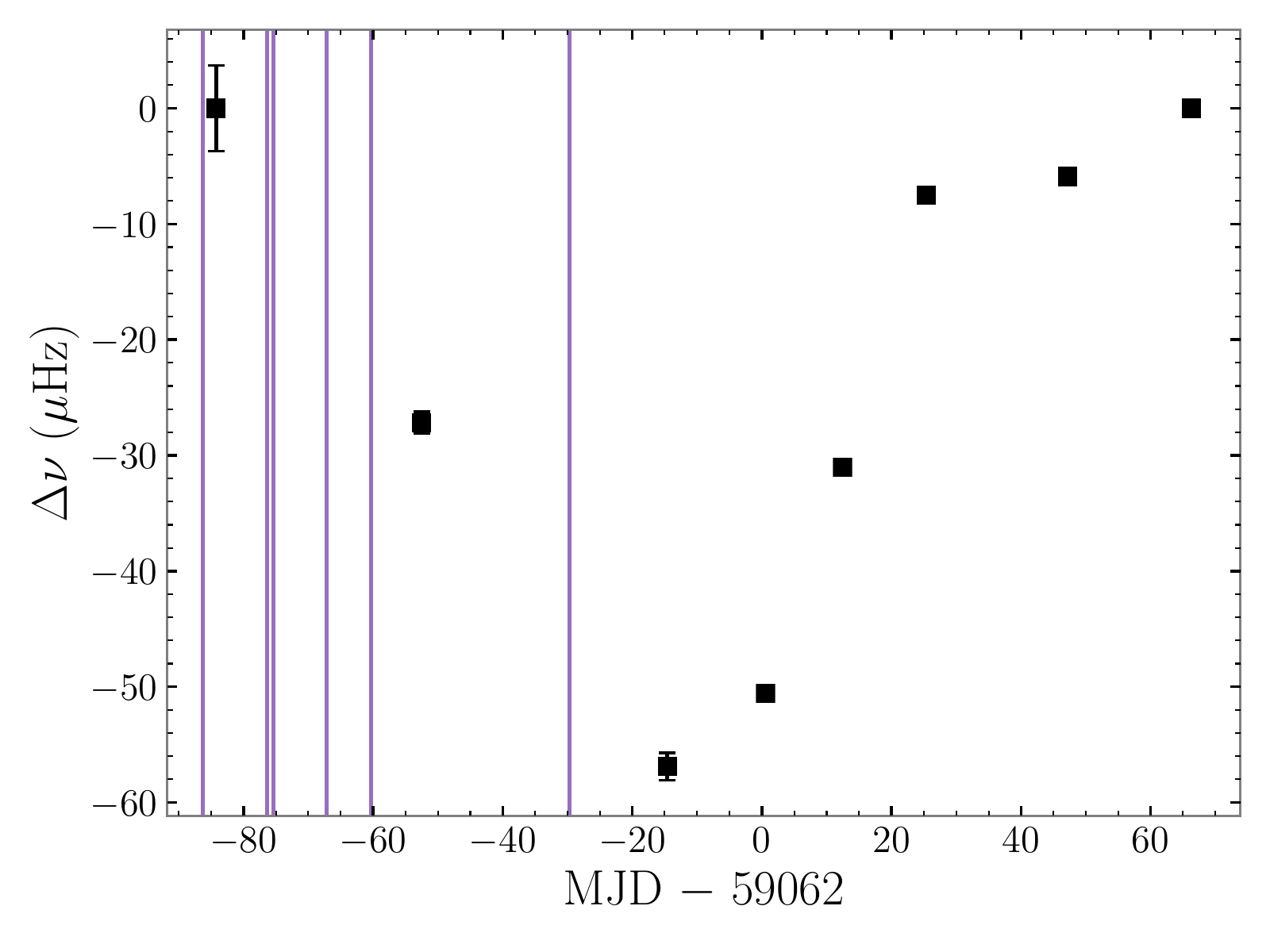}
    \caption{Variations in the spin-frequency of Swift J1818.0$-$1607 over time. Vertical lines correspond to high-energy bursts detected by \textit{Swift} \citep{Barthelmy2020GCN, Gronwall2020GCN, Bernardini2020GCN}.}
    \label{fig:spin}
\end{figure}

If these variations in the pulse profile are associated with fluctuations in the magnetospheric currents, then we might expect there to be some correlation with the spin-down behaviour or high-energy activity of the magnetar.
To place the emergent phenomena in context, we have plotted the inferred change in spin-frequency measured at each epoch (referenced to the spin-frequency on MJD 58977) after subtracting off a constant spin-down rate of $-2.37 \times 10^{-12}$\,s$^{-2}$ in Figure~\ref{fig:spin}. 
Also shown are epochs where high-energy bursts were detected by the Burst Alert Telescope on board \textit{Swift}. 
Visually, it is evident that at least two variations in the spin-down rate have occurred over the timespan covered by our observations, as indicated by the relatively sharp changes in $\Delta\nu$. 
The upward trend between MJD 59047 and 59087 could be a result of decreased particle outflows following the stabilisation of the inverted-spectrum component \citep[and the associated magnetospheric currents; see e.g.][]{Kramer2006}, while the flattening from MJD 59087 to 59128 could be associated with the evolution and eventual overlapping of the steep-spectrum component into the inverted-spectrum component.
Alternatively, these spin-frequency variations could be a result of the magnetar changing spin-down modes similar to what was reported by \citet{Champion2020}.
The GCN describing the hard X-ray/gamma-ray burst on MJD 59032 reported a peak count rate of $\sim$3000\,counts/\,s$^{-1}$, twice that of the initial burst that led to the discovery of Swift J1818.0$-$1607 \citep{Evans2020GCN}.
Its possible the resulting magnetic field reconfiguration associated with this burst triggered the emergence of the inverted spectrum profile component, initially through the transient P- and M-mode switching that we detected on MJD 59047.
Given our relatively sparse observing cadence, we cannot confirm a causational relationship between these two events.
Facilities with high observation cadences may be able to confirm or rule out a potential association.

Measurements of the magnetic geometries of magnetars is useful for both comparing predictions of how their magnetic fields may evolve over long timescales \citep{Tauris1998, Vigano2013, Gourgouliatos2014}, and for understanding their outburst mechanism \citep{Perna2011, Rea2012, Li2016}.
However, only a handful of magnetars have had their magnetic geometries constrained through radio polarimetry and fitting of their X-ray profiles, and various arguments have been made \citep{Kramer2007, Camilo2008, Levin2012}. 
From our geometric fits to the PA of Swift J1818.0$-$1607, we inferred a magnetic and viewing geometry of $(\alpha, \zeta) = (112^{\circ}\,^{+7}_{-9}, 99^{\circ}\,^{+7}_{-10})$, indicating it is an orthogonal rotator. 
A similar geometry was also inferred from polarimetry of the prototypical radio-magnetar XTE J1810$-$197. 
\citet{Camilo2007b} found both nearly aligned ($\alpha \sim 4^{\circ}$ and $\beta \sim 4^{\circ}$) and close-to-orthogonal ($\alpha \sim 70^{\circ}$ and $\beta \sim 20$-$25^{\circ}$) RVM-fits were both consistent with the data, depending on whether or not an OPM jump was included for the PA swing across the inter-pulse.
However, \citet{Kramer2007} argued a single RVM was insufficient to simultaneously fit both the main and interpulse.
Instead, they found that two separate fits to the individual components returned a consistent $\zeta = 83^{\circ}$ despite having recovering different values of $\alpha$ and $\beta$ for the main-pulse ($\alpha \sim 44^{\circ}$, $\beta \sim 39^{\circ}$) and interpulse ($\alpha \sim 77^{\circ}$, $\beta \sim 6^{\circ}$). 
\citet{Perna2008} and \citet{Bernardini2011} obtained similar constraints on the angles the line-of-sight and X-ray hotspot pole make with the spin-axis when the magnetar was in its outburst and quiescent states.
The deviation of both XTE J1810$-$197 and Swift J1818.0$-$1607 from being aligned rotators adds further credence to the argument that their broad radio profiles must be due to emission originating at large heights within the magnetosphere. 
It also rules out the rapid magnetic and spin axes alignment hypothesis we put forward in \citet{Lower2020} as a possible explanation for the apparent young age of Swift J1818.0$-$1607 despite the lack of an obvious associated supernova remnant. 

A complication to our geometric interpretation is the flipped PA swing direction (negative gradient instead of positive) we detected on MJD 59062.
Naively we could interpret this phenomena as either radio emission originating from the antipodal magnetic pole or our line of sight having undergone a latitudinal crossing of the magnetic pole.
Under the RVM, emission from the antipodal pole of the neutron star would exhibit a PA swing with the opposite sign, something that has been observed in a handful of pulsars where emission from both poles are detected as a main pulse and an interpulse \citep[e.g.][]{Johnston2019b}.
Similarly, geodetic precession of the relativistic binary pulsar PSR J1906$+$0746 resulted in a sign flip of its PA swing as the magnetic pole crossed our line of sight \citep{Desvignes2019}.
While a flipping of the emission to the opposite magnetic pole of Swift J1818.0$-$1607 could in principle explain the flipped PA swing, the averaged total intensity profile and spectra remains almost identical to that seen during the previous observation, making this scenario unlikely as the magnetic field and current configurations would have to be identical at both polar caps. 
If the shape of Swift J1818.0$-$1607 deviates from spherical symmetry due to crustal or magnetic stresses the spin axis can become offset from the total angular momentum vector.
This would cause the spin-axis to undergo free precession about the total angular momentum vector, resulting in both an apparent latitudinal and longitudinal evolution of the magnetic axis over time \citep{Pines1974}. 
However, free precession also presents an unlikely explanation for the profile and geometric variations. 
The short precession timescale required to explain our data would introduce periodic spin-down variations that are not detected in the high cadence timing by \citet{Champion2020} and \citet{Hu2020}.
Also, if the first timing event reported by those two studies is a true spin-up glitch, then free-precession is even more unlikely as the presence of pinned vortices within the neutron star core would rapidly dampen any precession \citep{Shaham1977}.
Further weight against the emission flipping between poles and the free precession arguments comes from both assuming a static, unchanging magnetosphere, where the observed profile variations are purely due to changes in the viewing geometry, whereas we have shown the magnetic and viewing geometries remain largely unchanged.
Additionally, our assumption that Swift J1818.0$-$1607 has a predominately dipole magnetic field geometry may be incorrect.

There are numerous theoretical and observational studies throughout the literature that point to magnetars possessing dynamic magnetic fields, where non-axisymmetric field geometries, higher-order multipoles and closed magnetic loops are suggested play an important role in describing the observed phenomenology \citep{Thompson1993, Thompson2002, Beloborodov2009}.
While complex multipole fields are likely to be present close to the surfaces of most neutron stars, the success of the RVM in describing the PA swings we observe suggests a more simplistic field geometry is associated with the radio emitting region of Swift J1818.0$-$1607. 
NICER observations of Swift J1818.0$-$1607 by \citet{Hu2020} showed the X-ray profile exhibits an unusually high pulse fraction for a profile with only a single component and noted it would be difficult to reproduce with the canonical two antipodal hotspot model.
They suggested this may instead be evidence the pulsed X-ray emission originates from either a single distorted surface hotspot or a two-component hotspot with differing temperatures.
A possible framework for describing such a hotspot configuration is provided by the magnetar corona model of \citet{Beloborodov2007}, where the high-energy and radio emission originates from either the closed magnetic field loops or open field lines emerging from two sites (starspots) on the neutron star surface -- somewhat analogous to coronal loops in the solar magnetic field that link pairs of sunspots.
Assuming this coronal loop interpretation holds true for Swift J1818.0$-$1607, we can explain the flipping of the position angle swing detected on MJD 59062 as being due to highly intermittent switching of the emission region between a more active `primary' and less active `secondary' starspot. 
A similar hypothesis was put forward by \citet{Kramer2007} to explain the PA swing of XTE J1810$-$197 during its 2003 outburst, where their preferred, dual RVM-fits were speculated to be evidence of radio emission originating from two active poles within a global multipolar field. 
Interpreting our RVM-fits geometrically, the inferred values of $\alpha$ from the normal/anomalous PA swings would correspond to the latitudinal positions of the two starspots on the neutron star, with the more active primary starspot positioned at $\alpha = 113 \pm 7^{\circ}$ and the secondary starspot located at $\alpha = 82 \pm 9^{\circ}$.
Slight wobbles in the PA swings could be an indicator the distribution of magnetic field lines linking these two starspots is not uniform, while temporal variations in the polarisation profile could be due to a changing emission height and variable plasma flows along the coronal loop connecting the two starspots.
Independent constraints on the viewing and emission geometry from X-ray observations, combined with continued radio monitoring would enable further tests of this hypothesis.
Additionally, a simple comparison of the radio and X-ray profile alignment could test whether the radio emission originates from closed magnetic field lines above the hotspot or from open field lines at heights comparable to the light cylinder radius (see, e.g., \citealt{Camilo2007c} and \citealt{Gotthelf2019} for a discussion on the X-ray and radio profile alignment of XTE J1810$-$197).

\section{Summary and conclusion}~\label{sec:conc}

Our wide-band radio observations of Swift J1818.0$-$1607 have revealed the magnetar possesses highly active and dynamic magnetosphere following its 2020 outburst. 
This is highlighted by our detection of new profile components, and the appearance of transient emission and polarisation modes. 
We showed the post-outburst magnetic geometry remains stable across most of our observations, where variations in the linear PA and profile polarisation can potentially be ascribed to changes in the relative emission height over time.
The reversed PA swing observed on MJD 59062 appears to be an anomalous outlier among our observations, which we speculate may be evidence of the radio emission at this epoch having originated from an additional, co-located magnetic pole that is offset from the primary pole by $\sim 30^{\circ}$ in latitude.

Continued monitoring of Swift J1818.0$-$1607 at radio wavelengths will allow for its magnetospheric evolution to be tracked as the current outburst progresses.
This includes the detection of any new emission mode changing or deviations from the magnetic geometry that describes the majority of the data presented here. 
For instance, a series of high-cadence observations may be able to catch a transition from the normally positive sloping PA swing to the seemingly rare negative swing we observed on MJD 59062.
Such a detection, combined with independent geometric constraints from fitting the X-ray profile and phase resolved spectrum of Swift J1818.0$-$1607, would provide an independent test of the coronal loop hypothesis we proposed as a potential explanation for this phenomena.

%%%%%%%%%%%%%%%%%%%%%%%%%%%%%%%%%%%%%%%%%%%%%%%%%%
%%%%%%%%%%%%%%%%%%%%%%%%%%%%%%%%%%%%%%%%%%%%%%%%%%
\section*{Acknowledgements}

The Parkes radio telescope (\textit{Murriyang}) is part of the Australia Telescope National Facility which is funded by the Australian Government for operation as a National Facility managed by CSIRO.
We acknowledge the Wiradjuri people as the traditional owners of the Observatory site.
This work made use of the OzSTAR national HPC facility, which is funded by Swinburne University of Technology and the National Collaborative Research Infrastructure Strategy (NCRIS).
This work was supported by the Australian Research Council (ARC) Laureate Fellowship FL150100148 and the ARC Centre of Excellence CE170100004 (OzGrav).
MEL receives support from the Australian Government Research Training Program and CSIRO Astronomy and Space Science.
RMS is supported through ARC Future Fellowship FT190100155.
This work made use of NASA's Astrophysics Data Service and the Astronomer's Telegram, in addition to the {\sc numpy} \citep{numpy}, {\sc matplotlib} \citep{matplotlib}, {\sc cmasher} \citep{cmasher} and {\sc tempo2} \citep{Hobbs2006, Edwards2006} software packages.
We thank the anonymous referee for their helpful comments and suggestions.

%%%%%%%%%%%%%%%%%%%%%%%%%%%%%%%%%%%%%%%%%%%%%%%%%%
%%%%%%%%%%%%%%%%%%%%%%%%%%%%%%%%%%%%%%%%%%%%%%%%%%
\section*{Data availability}

The raw UWL data will be available to download via the CSIRO Data Acess Portal (\url{https://data.csiro.au}) following an 18-month proprietary period. 
Other data products are available upon reasonable request to the corresponding author.



\bibliographystyle{mnras}
\bibliography{main} 

%%%%%%%%%%%%%%%%%%%%%%%%%%%%%%%%%%%%%%%%%%%%%%%%%%

% Don't change these lines
\bsp	% typesetting comment
\label{lastpage}
\end{document}